\documentclass[aip,preprint]{revtex4-1}

\usepackage{graphicx}
\usepackage[english]{babel}
\usepackage[T1]{fontenc}
\usepackage[colorlinks,linkcolor=blue,citecolor=blue,urlcolor=blue]{hyperref}
\usepackage{amsmath,amssymb}

\begin{document}

\title{Electron inertia effects in 3D hybrid-kinetic collisionless plasma turbulence}

\author{Patricio A. Mu\~noz}
\affiliation{Max Planck Institute for Solar System Research, 37077 G\"ottingen, Germany}
\email[]{munozp@mps.mpg.de}
\author{Neeraj Jain}
\affiliation{Center for Astronomy and Astrophysics, Technical University Berlin, 10623 Berlin, Germany}
\author{Meisam Farzalipour Tabriz}
\author{Markus Rampp}
\affiliation{Max Planck Computing and Data Facility, 85748 Garching, Germany}
\author{J\"org B\"uchner}
\affiliation{Max Planck Institute for Solar System Research, 37077 G\"ottingen, Germany}
\affiliation{Center for Astronomy and Astrophysics, Technical University Berlin, 10623 Berlin, Germany}

\date{\today}

\begin{abstract}
%
The effects of the electron inertia on the current sheets that are formed out of kinetic
turbulence are relevant to understand the importance of coherent structures in turbulence
and the nature of turbulence at the dissipation scales.
We investigate this problem by carrying out 3D hybrid-kinetic Particle-in-Cell (PIC) simulations
of decaying kinetic turbulence with our CHIEF code.
The main distinguishing feature of this code is an implementation
of the electron inertia without approximations.
Our simulation results show that the electron inertia plays an important role in
regulating and limiting the largest values of current density
in both real and wavenumber Fourier space,
in particular near and, unexpectedly, even above electron scales.
In addition, the electric field associated to the electron inertia dominates most of the strongest current sheets.
The electron inertia is thus important to accurately describe the properties of
current sheets formed in turbulence at electron scales.

\end{abstract}


\maketitle 

\section{\label{sec:intro}Introduction}

Collisionless turbulence is ubiquitous in space and astrophysical plasmas such as the solar wind,
planetary magnetospheres and the interstellar medium.
This collisionless (also often called kinetic) turbulence is inherently different in many aspects 
to the more widely studied and understood fluid or magnetohydrodynamic (MHD) turbulence \citep{Bruno2013,Howes2015a}.
The differences mainly have to do with the effects associated with the kinetic nature of these plasmas.
Those effects play an important role, in particular, at the dissipation scales of the turbulence cascade.
How this dissipation occurs is one of the fundamental problems in space plasma physics
and despite decades of theoretical, observational and numerical research is still poorly understood.
But in recent years significant observational and numerical investigations have shed new light on this problem.
These include spacecraft missions, like the Magnetospheric Multiscale Mission (MMS) with its high-resolution measurements in the Earth's magnetosphere,
Parker Solar Probe investigating the solar wind acceleration region;
as well as high-performance computer simulations,
which are capable of resolving a large range of spatial and temporal scales \citep{Chen2017y,Verscharen2019c}.

Turbulence is closely related to magnetic reconnection in different ways \cite{Matthaeus2011,Karimabadi2013}.
Magnetic reconnection is the fundamental process in the Universe by which magnetic energy
is quickly transformed into particle energy,
often in an explosive way like in solar flares and in the Earth's magnetopause and magnetotail \citep{Treumann2013b}.
Magnetic reconnection preferentially occurs in current sheets with a thickness around ion kinetic scales.
Since a few decades, the formation of current sheets and magnetic reconnection through them
has been observed and analyzed via simulations of turbulence by using a variety of fluid and kinetic plasma models \citep{Matthaeus1986,Servidio2010,Franci2017,Cerri2017,Haggerty2017a,Jain2021}.
This process has also been confirmed by in-situ observations in space plasmas,
ranging from the turbulent magnetosheath to the solar wind \citep{Yordanova2016,Voros2017,Khabarova2015a}.

The very often used inertialess-electron hybrid-kinetic codes, where ions are modeled kinetically while electrons are modeled as a massless fluid, are essential tools to understand plasma turbulence \citep{Winske2023}.
This is because they can resolve the physics associated to ion-kinetic scales, i.e., the ion skin depth and ion gyroradius in space, and the inverse of the ion cyclotron frequency in time.
From the turbulence standpoint, one of the reasons for the importance of those scales is that the nature of the typical collisionless turbulence changes
when the energy cascades from scales larger to smaller than the ion-kinetic length scales in space and astrophysical plasmas.
This energy cascade is expressed by the spectrum of, e.g., magnetic field fluctuations in the wavenumber (or frequency) space.
Above ion-kinetic scales, the turbulence spectrum can be fitted by a power-law with a power-law exponent that can be predicted
by MHD theories of turbulence.
Below ion-kinetic scales, the power-law exponent of  the magnetic turbulence spectra suddenly changes,
reflecting a change in the turbulent nature of the plasma.
This phenomenon is commonly known as ``spectral break'' in the wavenumber or frequency space.
This break has been often observed in-situ in magnetospheric and solar wind plasmas
and it often depends on parameters such as the plasma-beta \citep{Leamon1998,Parashar2018}.
One possible theoretical explanation (among many others)
is because at those small scales turbulence is composed by a much larger variety of different plasma waves that depend on ion scales, like kinetic Alfv\'en waves, in comparison with the pure MHD/Alfv\'enic-like waves at larger (inertial) scales \citep{Leamon1999, Boldyrev2015}.
In addition, hybrid-kinetic models can correctly describe ion-scale current sheets formed in collisionless turbulence, which are typically observed in space plasmas.
Because of those facts, among other reasons, there has been a large number of publications about the role of
ion-kinetic scales in collisionless turbulence with hybrid-kinetic codes
\citep[see, e.g.,][and references therein]{Franci2017,Cerri2017,Cerri2019,Jain2021}.

One important caveat of the usual inertialess-electron hybrid-kinetic approach is that it neglects all electron effects,
which are clearly relevant for the correct description of turbulence at electron kinetic scales.
There are different kinds of electron effects which can be important for those processes.
Some of them, such as the electron Landau damping with its associated dissipation,
are purely electron-kinetic effects which can be only captured with a fully-kinetic approach
for both ion and electrons, or with advanced reduced kinetic models such as gyrokinetics \citep{Told2015}.
Other electron effects, potentially relevant for turbulence, are electron temperature anisotropy instabilities, eventually driving turbulence and generating waves~\citep[see, e.g.,][]{Gary2015}.
Temperature anisotropy instabilities result as a consequence of the temperature components evolving differently along different spatial directions, in contrast to a simple but often used in hybrid codes scalar pressure (or temperature). This can be taken into account by means of a hybrid code that implements an evolution equation for the full electron pressure tensor, which in addition also models non-gyrotropic effects (resulting from off-diagonal terms in the pressure tensor) \citep[see, e.g.,][]{Kuznetsova1998}.
Effects related to the electron Larmor/gyro-radius are also relevant. They manifest, for example, as finite Larmor radius effects and/or in the form of wave-particle interactions such as electron gyro-resonances which are relevant for maser mechanisms and particle acceleration/diffusion and scattering in turbulence~\citep{Omura2008}.
Finally, the consideration of a finite electron mass (more commonly called electron inertia), is perhaps one of the most important electron effects
since it sets a specific length scale, the electron skin depth, which might control another spectral break in collisionless space plasma turbulence at the smallest dissipation scales \citep{Alexandrova2009,Sahraoui2013a,Huang2014e,Comisso2022}.
As for current sheets and magnetic reconnection through them,
electron inertia has three main effects on them:
it provides a physical mechanism to break the frozen-in condition
in the current sheets (even more efficient under guide-field conditions, see \citet{Ricci2004,Hesse2004}),
it limits the current sheet thinning \citep{Azizabadi2021,Jain2022}
and
provides additional pathways to dissipate magnetic energy
at the smallest kinetic scales, via, e.g., turbulence generated by electron shear flow instabilities \citep{Jain2014c,Jain2015c}.
Among those effects, the breaking of the frozen-in condition is perhaps the most important one, since it enables the non-ideal electric field that is necessary for magnetic reconnection
and its associated fast magnetic energy release.
%
Current sheets with a thickness around the electron-skin depth have also been recently observed in the turbulent Earth's magnetosheath thanks to the high-resolution measurements of the MMS mission.
In those currents sheets magnetic reconnection events without ion-coupling have also been detected \citep{Phan2018}.

Although a fully kinetic plasma approach (among other plasma models that include electron kinetics)
takes the electron inertia into account, it could be hard to disentangle its effects from the large number of other complex kinetic interactions at electron scales.
That is why a hybrid-kinetic plasma approach that takes the electron inertia into account is the ideal tool to investigate those effects separately.
Even though some hybrid-kinetic codes have implemented this plasma model,
most of them have considered some approximations which might hinder their applicability for scenarios
where strong electron-scale gradients are expected, like in turbulence and reconnection.

In the following we briefly review some of those codes which we already discussed
in a more extended way in \citet{Munoz2018} and \citet{Jain2022}.

The early attempts of hybrid-kinetic codes with electron inertia used the Darwin approximation of the Maxwell equations with the electromagnetic potentials in a Helmholtz decomposition (transverse and longitudinal part)~\cite{Forslund1971,Hewett1978}. The electron inertia was considered as a term in the transverse (to the magnetic field) electron current density equation and an evolution equation for the electron temperature was utilized.
However, they are restricted at least in the sense that the electron inertia is not considered along the parallel (to the magnetic field) electron equation of motion, and therefore the electron cyclotron frequency is not properly modeled in that direction.
Most of the later hybrid codes with electron inertia used a formulation based on electric and magnetic fields.
For example, \citet{Swift1996} developed a global magnetospheric hybrid code with the electron inertia included
only as an additional term in the form of an electron polarization drift in the equation for the advancement of the magnetic field (Faraday's law). This neglects any other electron inertia effect in the generalized Ohm's law.
\citet{Shay1998} and \citet{Kuznetsova1998} developed hybrid codes with electron inertia based on generalized electromagnetic fields  $\hat{\vec{B}}$ and  $\hat{\vec{E}}$.
This means that the electromagnetic fields are obtained indirectly. First, an equation resembling the Faraday's equation for $\hat{\vec{B}}$ and $\hat{\vec{E}}$ is solved. Then the electromagnetic fields $\vec{E}$ and $\vec{B}$ are obtained from $\hat{\vec{B}}$ and $\hat{\vec{E}}$ by using different approximations that impact the reliability of those models at electron-scale gradients.
For example, the spatial and temporal gradients of the density at electrons scales were neglected, as were the time derivatives of the ion current and/or the electron inertia.
On the other hand, those models have features that CHIEF does not have, such as an evolution equation for the full electron pressure, and thus they were more appropriate to model, for example, magnetic reconnection in scenarios
where it is 
Different from all previous codes, \citet{Valentini2007} developed a hybrid-kinetic code based on an Eulerian Vlasov formulation. Compared with the kinetic PIC formulation, this approach has the advantage of a much lower level of numerical noise at the expense of a larger memory requirement in order to store the 6D grid in phase space.
The electron inertia was included in a different way, as part of a Helmholtz equation for the generalized electric field that depends on, among other quantities, ion fluids quantities. The latter is a consequence of the use of the CAM (current advance) method used to couple the Maxwell equations with the (ion) Vlasov equation. This involves the evaluation the ion distribution function to get its momenta (ion bulk flow and ion pressure tensor), which are then used to determine the electric field.
This code has been mainly applied to solar wind turbulence.
Finally, we also mention the code by \citet{Amano2014}. The main feature of this hybrid-kinetic code is its capability to model plasma regions with low density and vacuum regions.
The electron inertia is included in the generalized Ohm's law for the electric field, but neglecting many of the spatial gradients associated to the currents and densities at ion scales.
For a more detailed discussion about the abovementioned codes and others, please see \citet{Munoz2018} and \citet{Jain2022}. In particular, a detailed comparison of the different approximations for the electron inertia can be found in Section 2.1 of \citet{Munoz2018}.

We took a different approach to the problem of implementation of electron inertia in a hybrid-kinetic plasma model
by developing the hybrid-kinetic PIC (Particle-in-Cell) code CHIEF
that implements the electron inertia term of the generalized Ohm's law without approximations \citep{Munoz2018}.
We have efficiently parallelized this code and applied it to the problem of 2D collisionless turbulence,
in particular assessing the effects of the electron inertia \citep{Jain2022}.

In the present work we extend the work of \citet{Jain2022} from 2D to 3D collisionless turbulence.
We focus on the effects of the electron inertia on turbulence and the current sheets formed out of it.
Many numerical simulations, including our previous study, are performed in 2D configurations
in order to save computational resources, even though turbulence is intrinsically 3D.
For example, turbulence theories, confirmed by both fluid and kinetic simulations, predict that 2D simulations of this process miss several critical properties compared to their realistic 3D counterparts,
such as the dominant waves in the nonlinear energy transfer and modified wave physics
that are involved in the anisotropic turbulent energy cascade, resulting in a modified dissipation rate \citep{Howes2014b,Li2016bn,Franci2018b,Gary2020}.
Current-sheet properties are also expected to be different in 3D compared to their 2D counterparts,
as evidenced by the large number of numerical investigations of 3D magnetic reconnection. \citep{Buchner1997,Horiuchi1999a,Fujimoto2011a,Che2011,Daughton2011}

There is thus a clear need for a study of 3D collisionless turbulence with self-generated current sheets and a focus on the electron inertia effects.
Our hypothesis is that the turbulence properties should exhibit a change of behavior
near the electron skin depth, and that the electron inertia should be relevant
for most of the current sheets formed in this turbulence.
To the best of our knowledge, this is the first attempt to address this problem with a hybrid-PIC kinetic plasma model that includes electron inertia.
Note that there have been previous 3D hybrid-kinetic simulations of turbulence with codes that include electron inertia,
but they have not focused on its effects and they were based on a hybrid-Vlasov approach \citep{Cerri2017a,Cerri2019,Sisti2021}.
Other publications have shown results of 3D fully kinetic PIC simulations of collisionless turbulence,
a model that intrinsically includes the electron inertia, with a focus on magnetic reconnection \citep{Rueda2021,Franci2022}.
This approach, however, does not allow to easily disentangle the pure electron inertia effects
from other presumably more relevant effects for the turbulent cascade, such as the electron Landau damping.
The same issue appears in some works comparing the turbulence properties of a variety of different plasma models,
including some with electron inertia \citep{Groselj2017,Gonzalez2019a}.
%
It is thus desirable to analyze turbulence simulations carried out
with a plasma approach where the pure electron inertia
effects can be switched on and off, and independently from other plasma effects.
This is in the spirit of comparing different plasma models, which is very valuable to provide new insights by isolating individual physical effects that cannot be easily disentangled in more complete simulation models carried out with, e.g., a fully kinetic model, even though they have full information about electron effects.
This approach has already been used in different contexts.
For example, \citet{Groselj2017} compared 2D turbulence simulations by using a hybrid-kinetic plasma model without electron inertia, a gyrokinetic code, and a fully kinetic one, among others.
They concluded that the electron Landau damping was responsible for the differences in the spectra between the hybrid and the other two plasma models, and a comparison between the gyrokinetic model with the fully kinetic and hybrid models revealed the importance of fast magnetosonic and Bernstein waves in the turbulent spectra (those two waves modes are ordered out in gyrokinetics).
Similarly, in the present article we shall compare two different plasma models in order to reach a conclusion on the
contribution of the electron inertia to the current sheets formed out of turbulence.
%

The remainder of this article is as follows. In Section \ref{sec:model} we describe the simulation model. The specific simulation setup is presented in Section \ref{sec:setup}. 
Section \ref{sec:results} shows our results. In particular, in Section \ref{sec:evolution} we show the time evolution of turbulence, useful to identify at which time the analysis are performed. Section \ref{sec:3d} describes the three-dimensional structure of turbulence, so to identify the spatial structure of current sheets.
The core of our new results about electron inertia effects is presented in Section \ref{sec:inertia}. We discuss our results and state our conclusions in Section \ref{sec:conclusion}. In Appendix \ref{sec:performance} we show the parallel scalability of our simulations with the newest version of our code, while in Appendix \ref{sec:energy} we show the variations in the total energy and its components in our simulations.

\section{Model and simulation setup \label{sec:model_and_setup}}

\subsection{Hybrid-kinetic plasma model \label{sec:model}}
We carry out 3D simulations of decaying turbulence by employing the hybrid-PIC code CHIEF \citep{Munoz2018}.
The hybrid-kinetic model with electron inertia that is implemented in this code has been described in detail in our previous publications \citep{Munoz2018,Jain2022, Jain2023}.
Here, we briefly recapitulate the main characteristics of the model:

We model ions kinetically via the Particle-in-Cell (PIC) method, as Lagrangian macro-particles whose distribution function $f_i$ obeys an ion Vlasov equation.
Via an averaging procedure in phase-space we obtain equations of motion for those macroparticles, formally identical to the Newton's equation of motion under the influence of electromagnetic fields,
but with the difference that those electromagnetic fields are determined by integrating (interpolating) the full electromagnetic fields by the macroparticles' shape function.
The Boris pusher is used to advance those macroparticles in time.
The same shape function is used, in turn, to determine the current density (and plasma density) at each grid point from the particles' position via $\vec{J}_i=e\int \vec{v}_i f_id^3\vec{v}$ and  $n_i=\int f_id^3\vec{v}$.

The electrons are modeled as a fluid employing the quasineutrality approximation $n_i=n_e$.
This implies that no continuity equation for the electron density is used, but rather
their density is equal to the ion density, which is directly determined from the particle information.
The quasineutrality approximation neglects charge separation and therefore limits our hybrid model
to phenomena with a frequency lower than the electron plasma frequency.
The electron momentum equation can be rewritten as a generalized Ohm's law for the electric field:
\begin{equation} \label{eq:e_ohm}
	\vec{E} = -\vec{u}_e\times \vec{B} -\frac{1}{en_e}\frac{\partial P_{e,jk}}{\partial x_k} - \frac{m_e}{e}\left(\frac{\partial \vec{u}_e}{\partial t}+(\vec{u}_e\cdot\vec{\nabla})\vec{u}_e\right) + \eta \vec{J}
\end{equation}
Here, the electron inertia effects are contained in the term proportional to $m_e$, which can also be written as $-\frac{m_e}{e}\frac{du_{e,z}}{dt}$.
$\vec{u}_e$ is the electron fluid velocity,
so that $(\vec{u}_e\times\vec{B})_z$ is the convective electric field, $\vec{E} + (\vec{u}_e\times\vec{B})$ the non-ideal electric field.
$P_{e,jk}$ is the $jk$ component of the electron pressure tensor. The repeated index $k$ implies a summation, so that $\partial P_{e,jk}/\partial x_k$ represents the $j=x,y,z$ component.
For the sake of simplicity, we used a scalar pressure and a simple isothermal equation of state here so that $p_e = \frac{1}{3}P_{e,kk} = n_ek_BT_e$, with $k_B$ the Boltzmann constant and $T_e$ the electron temperature, constant in time.
In Eq.~\eqref{eq:e_ohm}, $\eta \vec{J}$ is the resistive term.
We set $\eta$, the physical resistivity, equal to zero, but note that there is always
some numerical resistivity in the simulations that acts like an additional effective $\eta$ in the resistive term.
This is an effective residual resistivity that arises from the numerical errors of the discretization schemes in the algorithm itself, and is therefore hard to quantify and control.
This way, Eq.~\eqref{eq:e_ohm} shows that the  non-ideal electric field in our model can be supported by the resistivity or the
electron inertia term. In hybrid-kinetic codes without electron inertia, the non-ideal electric
field is only supported by collisional resistivity (either physical or numerical).
The mechanisms that support the non-ideal electric field 
are relevant because they are responsible for breaking the frozen-in condition and allowing magnetic reconnection in current sheets.
In collisionless space plasmas the resistivity is not very important,
so our model allows for a physical mechanism,
the electron inertia,
which is appropriate for those conditions and associated
to a length scale depending on plasma parameters, the electron inertia.

The electric and magnetic fields obey the Faraday and Amp\`ere's equation without displacement current,
which is an appropriate approximation for the considered frequency range of this plasma model: lower than the
electron plasma frequency but up to frequencies comparable with the electron cyclotron frequency, thus eliminating light wave propagation.
Note that most space plasmas are characterized by an electron plasma frequency which is much larger than the electron cyclotron frequency, making the assumption of both quasineutrality and range of applicability of the hybrid-kinetic plasma model with electron inertia (near and below the electron cyclotron frequency) valid.
By combining Eq.~\eqref{eq:e_ohm} with the Faraday's law, we obtain the following evolution equation for the magnetic field in the form of a generalized continuity equation
\begin{equation} \label{eq:curl_emom}
	\frac{\partial \overrightarrow{W}}{\partial t}
	   = \vec{\nabla}\times\left [\vec{u}_e\times \overrightarrow{W}\right]-\vec{\nabla}\times\left(\frac{\vec{\nabla} p_e}{m_en_e}\right),
\end{equation}
where
\begin{equation}\label{eq:generalized_vorticity}
	\overrightarrow{W}=\vec{\nabla}\times\vec{u}_e-e\vec{B}/m_e,
\end{equation}
is the generalized vorticity. Eq.~\eqref{eq:curl_emom} is solved by a parallel flux-corrected transport algorithm.
After the codes solves Eq.~\eqref{eq:curl_emom} for the generalized vorticity $\overrightarrow{W}$,
the magnetic field is calculated by combining the definition of  $\overrightarrow{W}$ in Eq.~\eqref{eq:generalized_vorticity} with Amp\`ere's equation in the following way
\begin{equation}\label{eq:elliptic_b}
	\frac{1}{\mu_0e}\vec{\nabla}\times\left(\frac{\vec{\nabla}\times\vec{B}}{n_e}\right)+\frac{e\vec{B}}{m_e} = \vec{\nabla}\times\vec{u}_i-\overrightarrow{W}.
\end{equation}
where $\vec{u}_i$ is the ion bulk flow velocity.
This is an elliptic equation that is solved by a parallel multigrid method.
The electric field is then calculated from the generalized Ohm's law Eq.~\eqref{eq:e_ohm}
by explicitly evaluating the time derivative term $\partial\vec{u}_e/\partial t$.
Note that, different from previous hybrid plasma approaches, our code CHIEF solves the previous equations without making any approximation for the calculation of the electron inertia terms (those proportional to $m_e$).
Further details are provided in \citet{Munoz2018,Jain2022, Jain2023}.
In particular, the current numerical implementation of this plasma model in our code CHIEF is described in detail in \citet{Jain2022}. 
That publication also describes the parallelization and scaling performance of its (quasi-)2D turbulence simulations.
For comparison purposes and to show the scaling behavior of our 3D CHIEF simulations, we discuss those points in the Appendix \ref{sec:performance}.

\subsection{Simulation setup \label{sec:setup}}
We use similar initial conditions as those often employed
in several previous studies of collisionless plasma turbulence mentioned in Section \ref{sec:intro}.
In particular we use a setup similar to the 2D one used by \citet{Jain2022}.
Note that, following the terminology of \citet{Howes2014b}, the (quasi-) 2D setup of \citet{Jain2022} is
the 2D perpendicular limit of our 3D setup, in the sense that there is no component of the allowed
wavevector parallel to the background magnetic field, but a wavevector spanning two dimensions on the plane perpendicular 
to the background magnetic field.
%
%
More specifically, in \citet{Jain2022} the initial fluctuations are imposed and current sheets form on the $y$-$z$ plane with a background magnetic field along the $x$-direction, $\vec{B}=B_0\hat{x}$.
We, on the other hand, use here a full 3D setup with the initial fluctuations imposed 
on the $x$-$y$ plane with a background magnetic field along the $z$-direction, $\vec{B}=B_0\hat{z}$. Consequently, the reversals of magnetic field mainly occur on the $x$-$y$ plane, which is associated to current sheets with their width on that plane and with their currents mainly along the $z$-direction.

This way, the initial magnetic field perturbations, $\delta \vec{B}_{\perp}$ (with the $\perp$ stands for the perpendicular direction to the background magnetic field), are the following randomly-phased, uncorrelated Alfv\'enic-like fluctuations,
\begin{equation}\label{eq:initial_perturbations}
  \delta \vec{B}_{\perp}=\sum_{k_x,k_y,k_z}\sqrt{\frac{2}{N_{k_x}N_{k_y}N_{k_z}}}\frac{\delta B_{rms}}{k}(k_y\hat{x} - k_x\hat{y}) \cos(k_x x + k_y y + k_z z+\phi_B(k_x,k_y,k_z)),
\end{equation}
where $\vec{k}$ is the wavevector of the fluctuations, $\phi_B$ is a random phase, $N_{k_x}$,$N_{k_y}$,$N_{k_z}$ are the number of excited modes
along each spatial direction, $B_{rms}$ is the root mean square of the magnetic field fluctuation amplitude, chosen as
$B_{rms}/B_0=0.24$,
which is a value commonly measured in the solar wind \citep{Bale2009}.
Note that the factor in front of the cosine term in Eq.~\eqref{eq:initial_perturbations} was chosen in order for each mode to have the same amplitude, i.e, $B_{\perp}(k_x,k_y,k_z)={\rm constant}$.
We excited the longest two wavemodes that can fit in the simulation box and all their combinations along the different directions, i.e,
$k_x=mk_0$ with $m=1,2$ and $k_0=0.12d_i$ ($d_i$ is the ion skin depth), and similarly for $k_y$ and $k_z$.
The ion bulk flow velocities have a similar form as in Eq.~\eqref{eq:initial_perturbations} but with different random phases $\phi_V$, so that they are uncorrelated with respect to the magnetic-field perturbations.
Their amplitudes are such that the total energy of the ion bulk flow velocity perturbations is the same as the total energy of the magnetic field perturbations, i.e.,
\begin{equation}\label{eq:initial_perturbations_v}
  \delta \vec{V}_{\perp}=\frac{V_A}{B_0}\sum_{k_x,k_y,k_z}\sqrt{\frac{2}{N_{k_x}N_{k_y}N_{k_z}}}\frac{\delta B_{rms}}{k}(k_y\hat{x} - k_x\hat{y}) \cos(k_x x + k_y y + k_z z+\phi_V(k_x,k_y,k_z)),
\end{equation}
where $V_A$ is the Alfvén speed based on the background magnetic field $B_0$.
Those initial magnetic field and ion bulk flow velocity perturbations at large (inertial) scales will cascade down to small kinetic scales, eventually forming current sheets.

Other physical parameters are: the background density is constant and equal to $n_0$.
The electron and ion plasma beta are $\beta_e=\beta_i=2\mu_0n_0k_B T_e/B_0^2=0.1$ with the (scalar) electron temperature $T_e$ equal to the ion temperature $T_i$.
No initial ion temperature anisotropies were considered.
The ion-to-electron mass ratio is $m_i/m_e=25$, which leads to a scale separation
between the ion and electron skin depths of $d_i/d_e=5$.
The ion gyroradius is defined as $\rho_i=v_{th,i}/\Omega_{ci}$,
where the ion thermal speed reads $v_{th,i}=\sqrt{k_BT_i/m_i}$ (with $k_B$ the Boltzmann constant)
and $\Omega_{ci}$ is the ion cyclotron frequency calculated on the background magnetic field $B_0$.
A similar definition holds for the electron gyroradius $\rho_e$, replacing ion by electron quantities.
This definition leads to a ratio between the ion skin depth and the ion gyroradius of $d_i/\rho_i=4.47$, a ratio between the ion gyroradius and electron skin depth of $\rho_i/ d_e = 1.12$
and $\rho_e/ d_e = 0.22$.

The numerical parameters can be summarized as follows.
The simulation box spans $N_x \times N_y \times N_z = 256 \times 256 \times 256$ grid points representing a physical size of $L_x \times L_y \times L_z = 51.2 d_i^3$.
This implies a grid cell size $\Delta x=\Delta y=\Delta z=1.0d_e$.
The timestep is chosen to be $\Delta t=0.004\,\Omega_{ci}^{-1}$.
We initially impose a homogeneous distribution of 60 protons per numerical cell.
The protons initially have a bulk drift speed given by Eq.~\eqref{eq:initial_perturbations_v}.
We use periodic boundary conditions in all spatial dimensions.

The choice of numerical parameters above (as well as the numerical algorithm itself), in particular
the number of particles per cell, can affect the energy conservation of the system, possibly influencing the physical results.
One way to improve energy conservation by reducing numerical noise is the use of higher-order shape functions,
like the second-order one that was used in our simulations.
In order to verify the conservation properties,
we show in Appendix \ref{sec:energy} the evolution of the total energy and their components of the simulation \textit{Run 1}, as well as the calculation method.
In general the total energy is conserved relatively well within 1$\%$ at the time of the diagnostics. Later there are larger deviations up to 4$\%$ towards the end of the simulation, as a result of the dissipation of turbulence.

In addition to the above described main 3D simulation (from now on denoted \textit{Run 1}) we carry out a supplementary simulation in order to analyze the electron inertia. This way our simulations are:

\begin{itemize}
\item \textit{Run 1}: The above-described main 3D hybrid-kinetic simulation with electron inertia ($m_e\neq 0$).
\item \textit{Run 2}: Same as \textit{Run 1} but without electron inertia, i.e., the equations that CHIEF solves are the same as those employed by the inertialess-electrons hybrid-kinetic codes ($m_e=0$).
\end{itemize}

\section{Results \label{sec:results}}


%
\subsection{Time evolution\label{sec:evolution}}

\begin{figure}[!ht]
	\begin{minipage}[b]{0.49\textwidth}
		\includegraphics[width=0.99\textwidth]{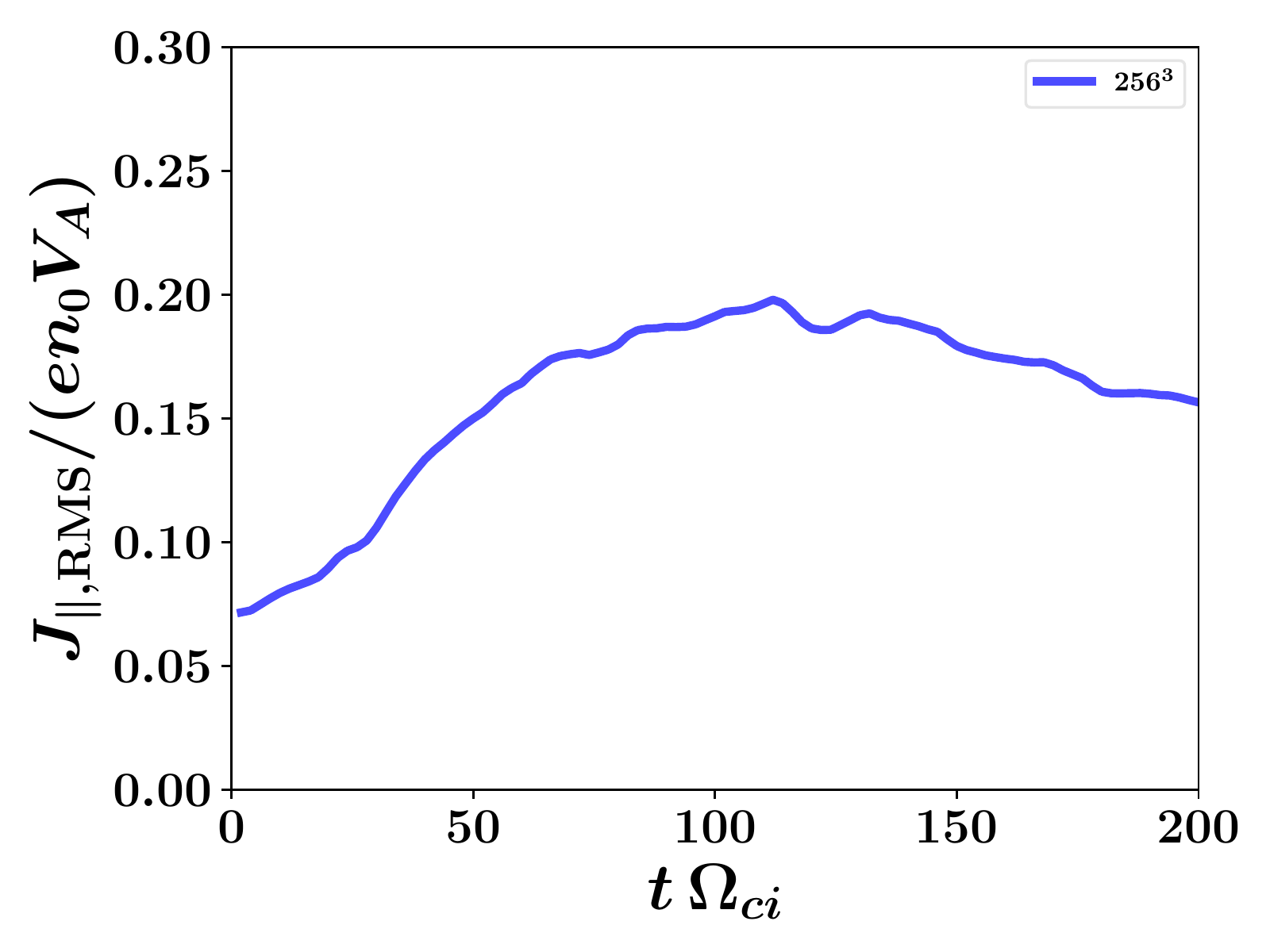}
	\end{minipage}
  \caption{Time evolution of the RMS values of the $J_{rms,\parallel}\sim J_z$ component of the current density
  for the standard 3D case (\textit{Run 1}, blue curve).
  \label{fig:std_jz}}
\end{figure}

As is customary in collisionless plasma turbulence simulations,
a meaningful way to quantify the evolution of the system are the RMS values of the current density.
This quantity is associated with the development of current sheets due to the turbulence as it cascades down to ion scales.
Fig.~\ref{fig:std_jz} shows the relevant component of the current density $J_{rms,\parallel}\sim J_z$ (parallel to the background magnetic field) for the standard \textit{Run 1}.
The 3D case without electron inertia (\textit{Run 2}) shows an evolution which is very similar to \textit{Run 1} and is thus not displayed here.




$J_{rms,\parallel}$ for \textit{Run 1} peaks around $t\sim 120\Omega_{ci}^{-1}$.
At this time the current sheets are quite disrupted, possibly due to reconnection, the formation of secondary structures, and their mutual interactions (see Fig.~\ref{fig:jz}(a3)).
A relevant time for analyzing the current sheet structure is therefore before the peak but not too early when current sheets
are not fully formed.
Hence, we choose $t=60\Omega_{ci}^{-1}$ for all analyses in the following (see Fig.~\ref{fig:jz}(a1-a2)).

\subsection{3D structure \label{sec:3d}}

\begin{figure}[!ht]
\includegraphics[width=0.99\textwidth]{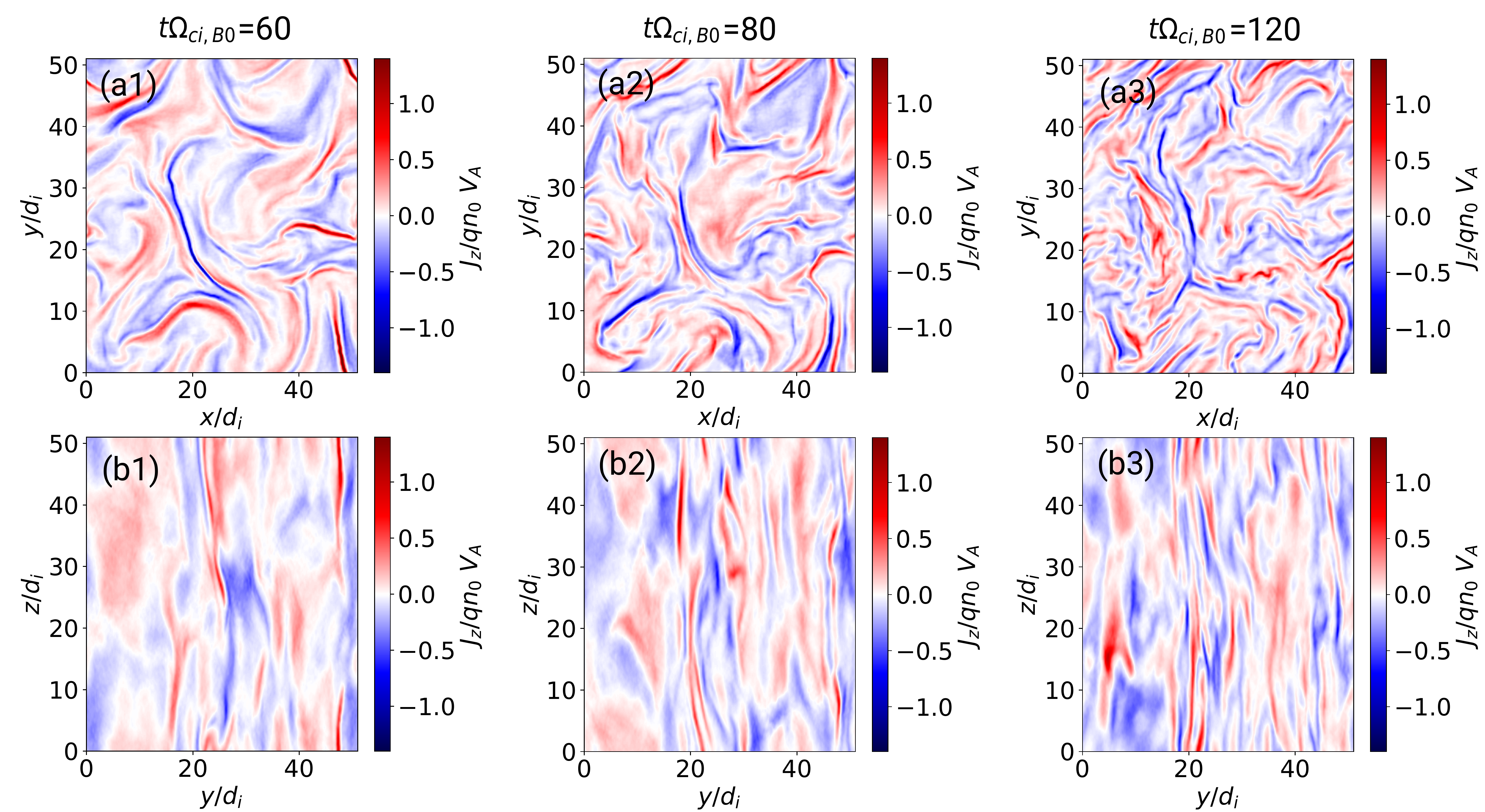}
        \caption{2D contour plots of $J_z$ for \textit{Run 1} at different times and planes.
        Top (a-) row: $x-y$ at $z=0$ plane.
        Bottom (b-) row: $y-z$ at $x=0$ plane.
        Left column: $\Omega_{ci}t=60$, middle column: $\Omega_{ci}t=80$, right column: $\Omega_{ci}t=120$.
        \label{fig:jz}
        }
\end{figure}

The top (a-)row of Fig.~\ref{fig:jz} shows 2D contour plots of the current density $J_z$ for \textit{Run 1} on a $x-y$ plane for selected times, namely the time of peak $J_{rms,\parallel}$, the chosen time for the analysis of current sheets $t=60\Omega_{ci}^{-1}$
and an intermediate time, $t=80\Omega_{ci}^{-1}$, for comparison.
The panels show the presence of a few current sheets (indicated by large positive and negative values in the current density $J_z$) with different
shapes and lengths, similar to the ones already found in previous simulations of collisionless plasma turbulence.
Most of those current sheets thin down up to typical length scales corresponding to the ion and electron skin depths.
The top (a-)row of Fig.~\ref{fig:jz} also demonstrates that most current sheets tend to get ``fragmented'' on the $x-y$ (reconnection) plane in the sense that their lengths get shorter with time.
This indicates that reconnection presumably already breaks the largest current sheets for times later than $t\sim (60-80)\Omega_{ci}^{-1}$.
The bottom (b-)row of Fig.~\ref{fig:jz} shows the structure of some of these current sheets along the parallel (background magnetic field) direction on a $y-z$ plane.
The sheets are elongated along the parallel direction and are possibly forming flux ropes,
in agreement with previous studies \citep{Sisti2021}.

\begin{figure}[!ht]
\includegraphics[width=0.99\textwidth]{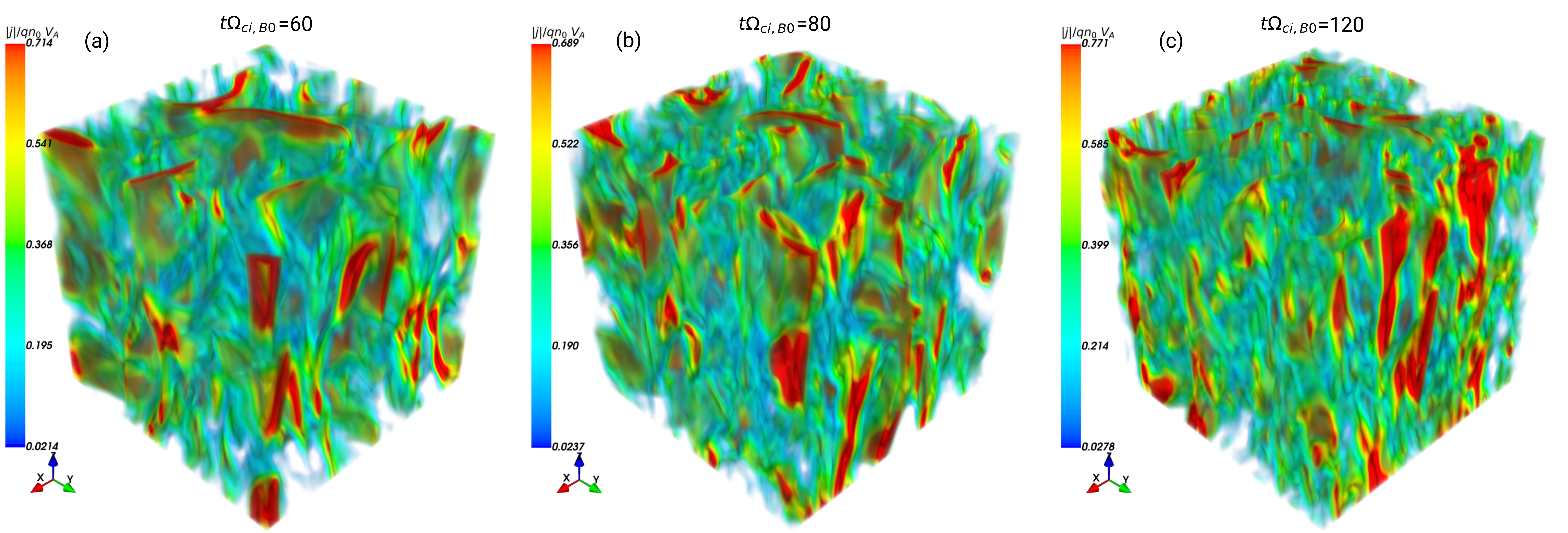}
        \caption{3D (volume-rendering) structure of $|J|$ at $\Omega_{ci}t=60$ (a), $\Omega_{ci}t=80$ (b), $\Omega_{ci}t=120$ (c). All for \textit{Run 1}. The spatial extension is 51.2 $d_i$ along each spatial direction (see also the 2D projections in Fig~\ref{fig:jz}).
        A movie showing the whole time evolution is also available (Multimedia available online).
        \label{fig:jz3d}
        }
\end{figure}

The three-dimensional structure of current sheets can be better distinguished in a volume-rendering of the absolute values of the total current density $|J|$ as shown in Fig.~\ref{fig:jz3d} (Multimedia available online).
This quantity turns out to be almost identical to $|J_{\parallel}|$,
which indicates that most of the current flows along the background magnetic field direction ($z$).
Note that the top faces of the cubic domain in Fig.~\ref{fig:jz3d}
correspond (bar absolute value) to the 2D plots shown in the top (a-)row of Fig.~\ref{fig:jz}, while the front right faces in Fig.~\ref{fig:jz3d}
correspond to the bottom (b-)row of Fig.~\ref{fig:jz}.

\subsection{Electron inertia effects \label{sec:inertia}}

In order to analyse the effects of the electron inertia term
on the current-sheet properties, we first
calculate the one-dimensional integrated (reduced) Fourier spectral power
of the current density.
This is done, in general,
by integrating the power of a given scalar field $A$ in the wavenumber Fourier space, $A^2(k_x,k_y,k_z)$,
on each concentric non-overlapping shell in the perpendicular wavenumber $k_{\perp}\approx k_x\hat{x}+k_y\hat{y}$
with a width of $\Delta k=2\pi/L_x$.
This is followed by either an integration along $z$ (parallel direction) in order to get the
averaged power as a function of the perpendicular wavenumber, $P(k_{\perp})$,
or an integration along all those shells in $k_{\perp}$-space
in order to get the averaged power as a function of the parallel wavenumber, $P(k_{\parallel})$.

\begin{figure}[!ht]
	\includegraphics[width=0.99\textwidth]{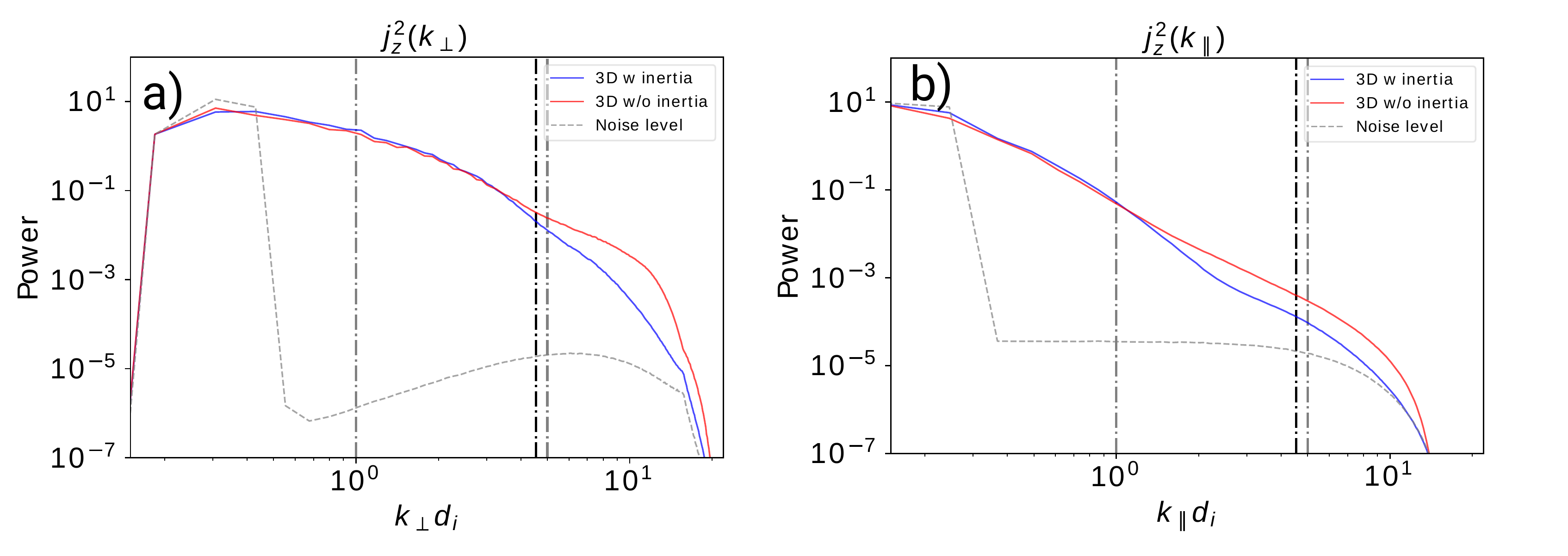}
        \caption{1D integrated (reduced) spectral power of the parallel current density fluctuations $J_z$ for the 3D cases (\textit{Run 1}, blue curve and \textit{Run 2}, red curve)
         at $\Omega_{ci}t=60$.
        (a) $P(k_{\perp})$. (b) panel:  $P(k_{\parallel})$.
        The gray curve is the initial spectral power of \textit{Run 1}, representing the noise level.
        See the main text for details of the calculations.
		The left dashed vertical gray line is drawn at $k=d_i^{-1}$, while the right  dashed vertical gray line at $k=d_e^{-1}$.
		The black dashed vertical black line is located at $k=\rho_i^{-1}$.
		\label{fig:jspectra}
        }
\end{figure}

The results of this calculation for the parallel current density $J_z$ (associated with the current sheets) are shown in
Figure~\ref{fig:jspectra}.
Although this is not a standard diagnostic for turbulence simulations
in comparison with the magnetic field spectra,
it helps to assess the scales at which the electron inertia effects
influence the current sheets.
Note also that the spectral features of the current density $J_z$ in Figure~\ref{fig:jspectra}
do not directly map to those of the magnetic field spectra $B_{\perp}$.
So it is not possible to conclude that features (or absence of them)
in the spectra $J_z$ should also appear in the spectra $B_{\perp}$.
For example, spectral power-law indices differ in different regimes, in particular above and below ion-scales (as often investigated in previous works analyzing ion spectral breaks), and the rapid drop at the dissipation scale in the spectra of $J_z$ does not appear in the spectra of $B_{\perp}$.
Figure~\ref{fig:jspectra}(a) shows that $P(k_{\perp})$ is very similar for
the two presented cases at ion scales, i.e., for wavenumbers smaller than those corresponding to the electron skin depth $k d_e=1$ (right gray dashed line).
Those spectra are well above the noise level, as shown by the gray curve which represents the initial spectral power of \textit{Run 1}. Note that the initial bump in the gray curve for low wavenumbers represents the initial Alfv\'enic wave perturbations.
For wavenumbers larger than $k d_e=1$, $P(k_{\perp})$  of the current density
is larger for the 3D case without electron inertia (\textit{Run 2}) than for the case with
electron inertia (\textit{Run 1}) around and below the wavenumber corresponding to the electron skin depth.
This implies a flatter spectrum above electron scales for the case without electron inertia.
Note that, coincidentally, the differences between \textit{Run 1} and \textit{Run 2} also
appear near the scales associated with the ion gyro-radius, $k \rho_i=1$,
since this quantity is close to the electron skin depth for our parameters.
However, the differences in the spectra cannot be interpreted as the result of effects related to the ion gyroradius because
they only appear when the electron inertia is disabled in \textit{Run 2}, so they must be related to the latter.
Figure~\ref{fig:jspectra}(b) for $P(k_{\parallel})$ shows a similar behavior as for $P(k_{\perp})$ but
with the important difference that
the electron inertia effects starts to play a role already at scales much larger than
the electron skin depth (approximately at $kd_i\sim 2$ or $kd_e\sim 0.4$).
This might imply that the current sheet structure along the parallel direction
significantly differs 
between the cases with and without electron inertia 
already at scales where electron effects are not normally expected to play a role.
Note that this can only be observed in a 3D turbulence geometry; the 2D-limit misses this effect, even though current sheets are also formed.


Note that the power spectrum evolves in time, so it is hard to make precise comparisons for different simulations.
Another caveat is that Figure~\ref{fig:jspectra} does not allow to pin down whether current sheets are the only responsible structures for the power at electron scales,
where differences appear. The following discussion, however, shall provide evidence that this is indeed the case.

\begin{figure}[!ht]
	\includegraphics[width=0.6\textwidth]{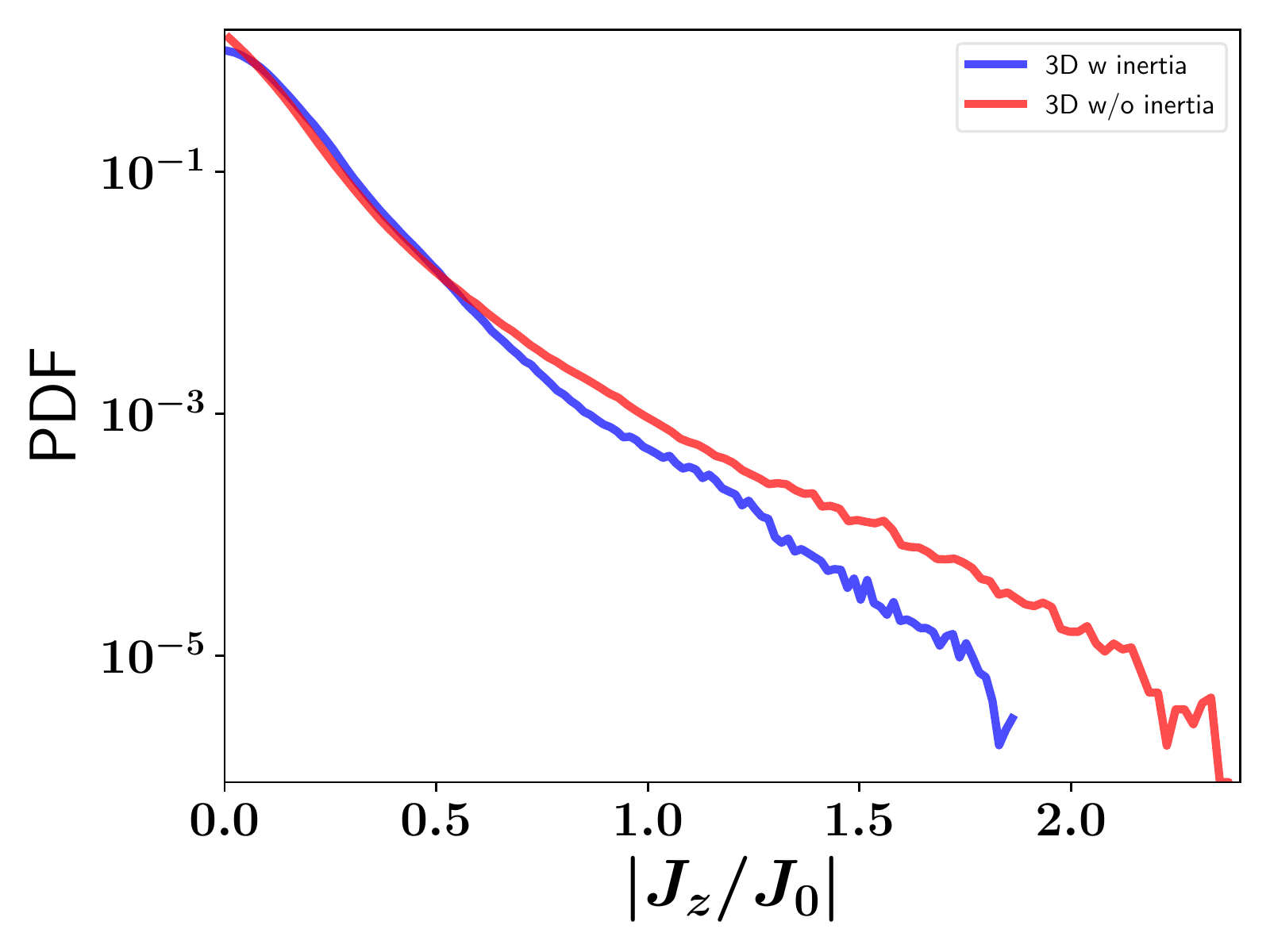}
        \caption{Histogram of values of the current density $J_z$ for the 3D cases (\textit{Run 1}, blue curve, and \textit{Run 2}, red curve) at $\Omega_{ci}t=60$.
        \label{fig:jhisto}
        }
\end{figure}

In order to identify where the main differences
caused by the electron inertia appear,
we analyze the values of $J_z$ in real space.
We note that the general excess of spectral power in the parallel current density fluctuations
of \textit{Run 2} (no inertia) vs \textit{Run 1} (inertia), cf. Fig.\ref{fig:jspectra},
is also present in real space.
Indeed, Fig.~\ref{fig:jhisto} shows the histogram of the values of $|J_z|$ at each grid point for the different simulations.
From this Figure we can see that the distribution of values of $|J_z|$ of the
simulation run without inertia (\textit{Run 2}) shows two main differences with respect to the case with inertia (\textit{Run 1}):
the maximum values are larger and
the volume occupied by grid points with a given value of $|J_z|$ ($\gtrsim0.5J_0)$ is also larger.
All in all, we conclude that the electron inertia effects tend to limit
the maximum values of parallel current density, as well as to increase the volume that is occupied by large values of $|J_z|$ ($\gtrsim0.5J_0)$.

\begin{figure}[!ht]
    \includegraphics[width=0.99\textwidth]{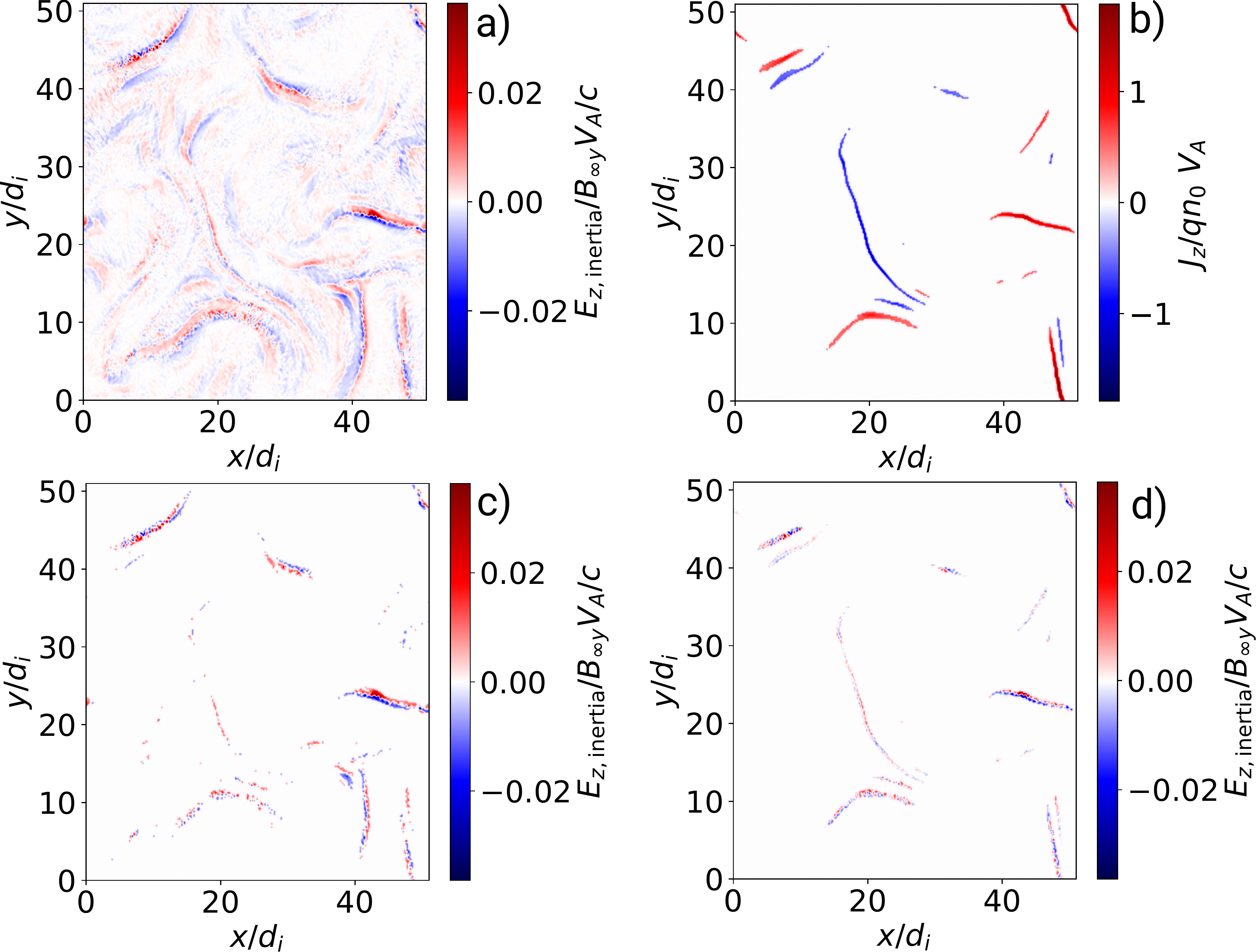}
         \caption{
         (a)  Contour plot of the electron inertia term $-\frac{m_e}{e}\frac{d\vec{V}_e}{dt}$.
         (b) Contour plot of values of $J_z>J_{z,RMS}$
         (c) Contour plot of values of $E_{z,inertia}>3E_{RMS,z,inertia}$
         (d) Values of $E_{z,inertia}$ at points where $J_z>3J_{z,RMS}$.
         All plots are done for \textit{Run 1} at $\Omega_{ci}t=60$ and on the $x-y$ plane at $z=0$.
         \label{fig:ez_inertia2}
        }
\end{figure}

The previous results indicate that the electron inertia affects the current-density strength,
but they do not reveal the causal effects.
In order to analyze these, we calculate the contribution of the electron inertia to the electric field in the generalized Ohm's law Eq.~\eqref{eq:e_ohm}.
We focus, in particular, on the component of Eq.~\eqref{eq:e_ohm} that is aligned with the magnetic field,
which in our setup is mainly $E_{\parallel}\approx E_z$:
\begin{equation}\label{eq:ohm2}
	E_z + (\vec{u}_e\times\vec{B})_z = -\frac{m_e}{e}\frac{du_{e,z}}{dt}
\end{equation}


Figure~\ref{fig:ez_inertia2}(a) shows the right hand side of Eq.~\ref{eq:ohm2},
the electron inertia term, for the standard \textit{Run 1}.
By comparing with Fig.~\ref{fig:jz}(a1),
we notice that the values of the electron inertia are the largest near some of the strongest current sheets.
That statement can be even better justified by the following procedure:
First, we identify the strongest current sheets by selecting
the grid points with a value of $J_z>3J_{z,RMS}$, i.e., three standard deviations over the root mean square value.
According to Fig.~\ref{fig:std_jz}, $J_{z,RMS}\approx 0.15J_0$ at the time of our diagnostics.
The results are shown in Fig.~\ref{fig:ez_inertia2}(b).
Note that this procedure does not allow the identification of all current sheets, since some of them may have current density strengths under the threshold. And this procedure could also pick up values that are not part of current sheets but only strong fluctuations at isolated points due to numerical noise.
Nevertheless, it is a simple and convenient first-order approximation for identifying the strongest current sheets.
The second step consists of selecting all the points of the electron inertia term plot in Fig.~\ref{fig:ez_inertia2}(a)
within three standard deviations of the RMS values of the same term ($3E_{RMS,z,inertia}$).
The results are shown in Fig.~\ref{fig:ez_inertia2}(c), which highlights the regions where the electron-inertia term is dominant.
The third and final step is selecting the values of the electron-inertia term that are located in the regions with strongest current sheets as per the first step, i.e.,
$J_z>3J_{z,RMS}$.
The results are shown in Fig.~\ref{fig:ez_inertia2}(d).
By comparing with the panel (c) we can thus infer that
the electron inertia is dominant in most of the strongest
current sheets.
But there are also regions with large values
of electron inertia that are not located
on current sheets, as well as
current sheets with values of electron inertia that are not particularly large.
Note that this method also has the caveat that it is for a particular 2D plane of the 3D simulation and for a specific time. Analysing data from different 2D planes and different times will lead to current sheets with a different distribution of electron inertia, but the aforementioned trend seems to hold in general.

\begin{figure}[!ht]
                \includegraphics[width=0.7\textwidth]{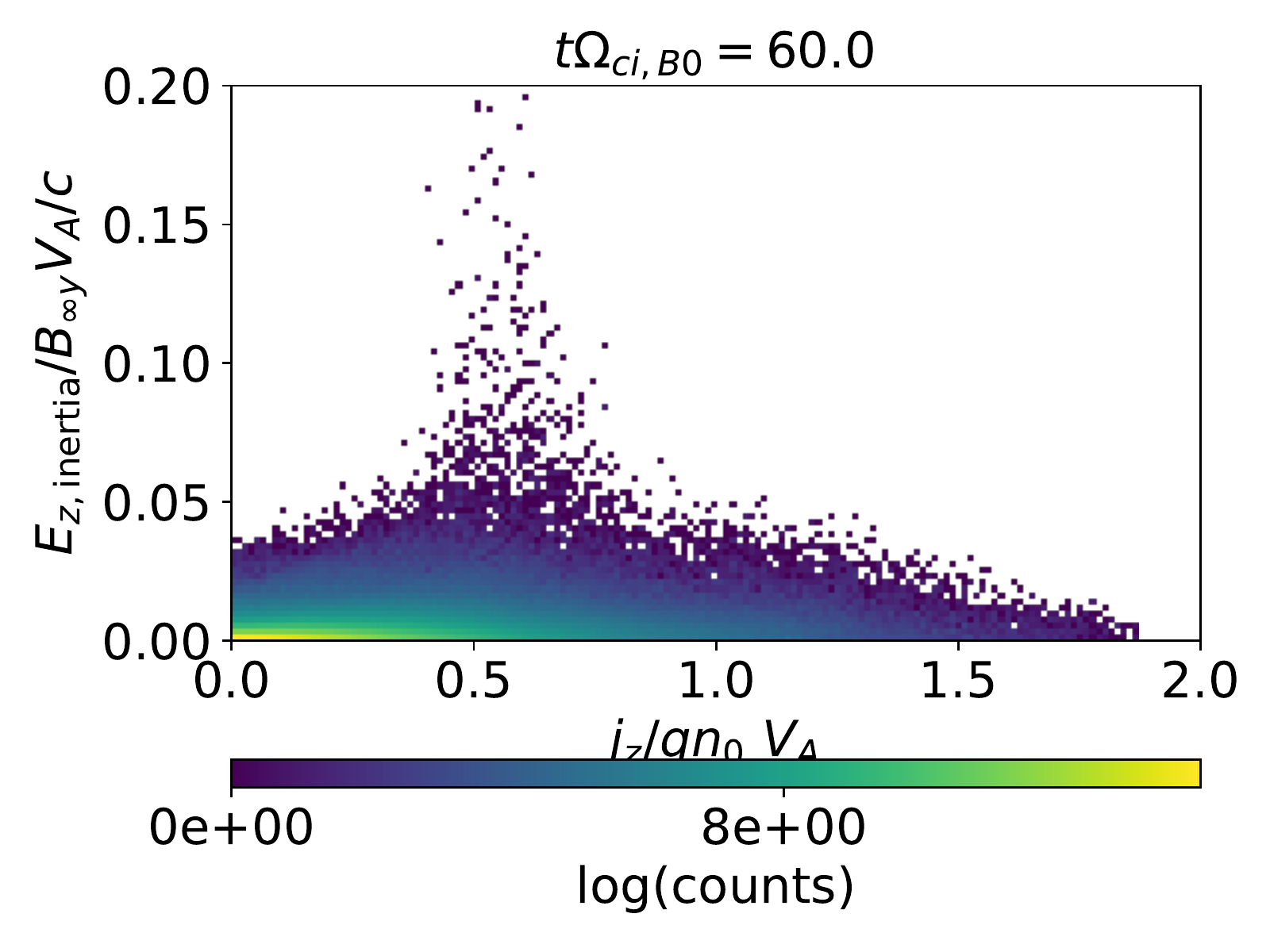}
        \caption{
        Histogram of the values of the electron inertia term versus the values of the current density $J_z$ at each grid point in the 3D simulation domain at $\Omega_{ci}t=60$, and for \textit{Run 1}.
        \label{fig:ez_inertia_histogram}
        }
\end{figure}

A different way to identify where the electron inertia becomes relevant is illustrated by Figure~\ref{fig:ez_inertia_histogram}, which shows the histogram of the distribution
of values of the parallel current density versus the electric field associated with the electron inertia.
The Figure shows a strong peak of $E_{z,inertia}$ near the (normalized) value 0.5 of the current density $J_z$, which are not the largest values of current density that are present in the simulation (see Fig.~\ref{fig:jhisto}).
This value is actually a bit larger than $3J_{z,RMS}$.
The explanation for this observation relates to the fact
that the maximum values of the electron inertia term are
located rather at the edges of the current sheets, with corresponding lower values of the current density.
This is a known property of the electric field associated with the electric inertia,
which has already been found by previous 2D laminar magnetic reconnection
simulations \citep{Hesse2004}.
Interestingly, the peak of electron inertia at $|J_{z}|/J_0>0.5$
correlates with the values of current density above which
the distribution of currents starts to diverge
between \textit{Run 1} and \textit{Run 2} (3D with and without electron inertia), see Fig.~\ref{fig:jhisto}.
This adds further evidence to our finding that electron inertia
plays an important role in sustaining and limiting the growth of most of the strongest current sheets, in particular in a realistic 3D geometry.




\section{Discussion and Conclusions}
\label{sec:conclusion}

We have carried out 3D collisionless turbulence simulations by using our CHIEF code,
which implements a hybrid-kinetic plasma model with a full consideration of the electron inertia term in the generalized Ohm's law.
In particular, we assessed the influence of the electron inertia on the current-sheet properties formed out of turbulence.
Due to diverse considerations discussed in the introduction,
the electron inertia should play an important role on those processes.
However, their effects have been
so far not investigated in detail in previous numerical investigations of plasma turbulence,
even though this phenomenon has been analyzed with plasma models that include those effects,
such as fully-kinetic models.
Our previous study has investigated this issue but within
a quasi-2D setup~\citep{Jain2022}. Here we take a step forward by analyzing a more realistic 3D
setup. It is known that turbulence is inherently 3D and the 2D limit misses
several important aspects, which could potentially affect current sheet properties.

Previous numerical studies of collisionless turbulence have been normally carried out
with plasma models that do not include electron inertia, such as MHD or
usual hybrid-kinetic models.
Those that do include electron inertia, such as
the fully kinetic or gyrokinetic plasma models, do not provide an easy
way to disentangle the pure electron inertia effects from other electron kinetic effects
such as Landau damping.
In this sense, the present article consistently compared two different plasma models, hybrid-kinetic with and without electron inertia.
This was done within the same numerical code,
in order to avoid uncertainties due to the use of different algorithms or numerical schemes.

Our results provide evidence that the effects of electron inertia modify the distribution of parallel current density (associated with current sheets)
near electron scales.
More specifically, the spectral power of current density
is overestimated if electron inertia is not taken into account.
This takes place near and below electron scales
for the distribution of current density in dependence on
the perpendicular wavevector,
while at scales larger than electron scales in dependence
on the parallel wavevector.
This means that the structure of current sheets along the background magnetic field
direction is significantly altered if the electron inertia is taken into account.
In other words, this modification does not only occurs at electron scales as expected,
but also above it, possibly in the range of ion scales.
This contradicts our intuitive expectation of electron effects being
limited to only electron inertial scales and will be further analyzed elsewhere. 
Note that this effect can only be observed in a 3D geometry; it is missed in the 2D limit.

Furthermore, we found that the electron inertia also plays an important role in
regulating and limiting the largest values of current density.
This leads to the formation of current sheets with
lower and presumably more physical values compared to plasma models
without electron inertia, since a plasma model including electron inertia
is more complete than one without it.


Finally, we showed that the electron inertia dominates
most of the current sheets.
In agreement with previous studies, the maximum values of the electric field associated with the electron inertia
are not reached at the peak values of current density, but rather at lower values.
Interestingly, those values are the same above which
the distribution of currents starts to diverge
between the cases with and without electron inertia.

Caveats of our results can be divided in those intrinsic to the model and those related to the parameter choice. We explained both next.

An important caveat intrinsic to the utilized plasma model is the consideration of
only the electron inertia as a physical mechanism that can generate a non-ideal electric field
and break the frozen-in condition, allowing magnetic reconnection.
There are other physical mechanisms that also allow this process;
the most important among them is arguably the contribution of the non-gyrotropic electron pressure tensor
to the non-ideal electric field.
This is neglected in our simulations with the CHIEF code. This mechanism
is related to a more accurate description of the electron thermodynamics,
which the CHIEF plasma model neglects by assuming a scalar electron temperature governed by an isothermal equation of state.
This is a common choice in hybrid-kinetic models and has been shown to reliably reproduce many features of observed space plasmas.
Technically, our model could easily adopt, for example, a polytropic equation of state if needed.
There are also advanced equations of state used in other hybrid codes aimed at modeling magnetic reconnection,
which have proved to reproduce many of the features of this process that can be observed within fully-kinetic plasma models~\citep{Le2016}.
A more accurate modeling of the electron thermodynamics could use an evolution for the electron pressure (either scalar or tensor), which involves setting additional constants, such as a viscosity \citep{Shay1998}.
If a tensor pressure is used, its evolution equation also needs to consider additional numerical constants
that have to be empirically determined, such as an isotropization term to account for
pitch angle scattering due to electron temperature anisotropy instabilities, which are not self-consistently modeled by that evolution equation \citep{Hesse1995}.

Our choice of parameters regarding the plasma-beta ($\beta_e=0.1$) actually also allows to partially justify the usage of the simple scalar equation of state:
this low-beta regime, where magnetic pressure dominates over the thermal pressure,  implies that the electron gyroradius is smaller than the electron skin depth ($\rho_e/ d_e = 0.22$).
This, in turn, implies that effects related to electron thermodynamics (e.g., equation of state, pressure tensor, associated to the electron gyroradius)
should be less important in comparison with the electron inertial effects (which are associated to the electron skin depth).
Therefore, a simple scalar equation of state is more appropriate in our case than in higher-beta plasmas.

Regarding other caveats related to numerical parameters, the probably most important one
is the choice of a reduced (compared to the real value) proton-to-electron mass ratio.
Our chosen value is $m_i/m_e=25$, which is of course not physically realistic but chosen due to computational resource limitations,
as usual in this kind of 3D kinetic plasma simulations.
This could lead to an overestimation of electron inertial effects, however, electron inertia alters the turbulence spectra at electron scales $kd_e =\sqrt{k_{\perp}^2+k_{||}^2}\,d_e\sim 1$ such that $k_{||} << k_{\perp}$. This is expected to be true for the real mass ratio as well.
In any case, our results provide the first evidence for the effects of the electron inertia in this system 
and represent a first step towards a better understanding of the electron inertia in current sheets formed in kinetic plasma turbulence.
Future scaling studies about the dependence of the results
on the mass ratio and other parameters have to be carried out in order to establish quantitative predictions which can be directly applicable to space plasmas.

The resolution of our simulations is also not specially high, but sufficient.
Indeed, note that the electron gyroradius, as a result of the low plasma beta,
is below the grid cell size, $\rho_e/\Delta x = 0.22$.
This quantity, however, does not need to be larger than the grid cell size
in our plasma model, which considers electrons as a fluid, and therefore does not include electron gyro-resonances or any related phenomena where the electron gyroradius plays a role.
On the other hand, the ion gyroradius needs to be resolved by the grid cell size to properly consider ion kinetic effects such as gyroresonances. In our case, it is resolved by $\rho_i/\Delta x = 1.12$.

Another different type of concern regarding our results is the level of numerical noise.
This mainly comes from the number of numerical particles per cell,
whose value was chosen to be a good compromise between the convergence of values of some physical quantities based on some previous test simulations and the limitations of our computational resources.
In order to reduce the numerical noise level,
we used a higher-order interpolation scheme/shape function,
which smooths out the current density at grid-scales, diminishing the level of numerical noise compared to the more usual linear schemes, and it can contribute to modify
the turbulence spectra at wavenumbers corresponding to the grid cell size.
In any case, the numerical noise does not affect much
our analyzed quantities (in particular current densities)
in comparison with other more sensitive parameters to the numerical noise,
such as temperatures or electric fields.
This is reflected in the low level of numerical noise
in comparison with the physical current density spectrum and
the good energy conservation properties displayed by our simulations.

In general, we conclude that the electron inertia
plays an important tole role to sustain and limit the current density of most of the strongest current sheets formed out of turbulence, in particular around electron scales and more so in a 3D geometry.
Unexpectedly, some effects of electron inertia also appear above electron scales,
possibly in the range of ion scales.
It is thus important to properly take the electron inertia into account in order
to better describe the current sheet properties formed out of collisionless turbulence.

\begin{acknowledgments}

The authors acknowledge the support by the German Science Foundation (DFG) 
projects JA 2680-2-1, MU-4255/1-1, BU 777-16-1 and BU 777-17-1. Computations were performed at the Max Planck Computing and Data Facility (MPCDF).
We thank the referees' comments who helped us to improve this paper.

\end{acknowledgments}

\section*{Author Declarations}

\subsection*{Conflict of Interest}

The authors have no conflicts to disclose.

\subsection*{Author Contributions}

\textbf{Patricio A. Mu\~noz:}
Conceptualization (supporting); 
Data curation (lead);
Formal analysis (lead); 
Funding acquisition (supporting); 
Investigation (lead); 
Methodology (equal); 
Software (supporting); 
Validation (lead); 
Visualization (lead); 
Writing – original draft (lead);
Writing – review \& editing (equal).
\textbf{Neeraj Jain:}
Conceptualization (supporting); 
Formal analysis (supporting); 
Funding acquisition (supporting); 
Methodology (equal); 
Software (supporting); 
Writing – review \& editing (equal).
\textbf{Meisam Farzalipour Tabriz:}
Data curation (supporting);
Software (lead); 
Validation (supporting); 
Visualization (supporting); 
Writing – review \& editing (equal).
\textbf{Markus Rampp:}
Project administration (supporting);
Resources (equal); 
Software (supporting); 
Supervision (supporting); 
Writing – review \& editing (equal).
\textbf{J\"org B\"uchner:}
Conceptualization (lead); 
Funding acquisition (lead); 
Project administration (lead);
Resources (equal); 
Supervision (lead); 
Writing – review \& editing (equal).

\section*{Data Availability Statement}

The data that support the findings of this study are available from the corresponding author upon reasonable request.

\appendix
\section{Code performance and parallel scalability
\label{sec:performance}}

In order to assess the parallel scalability of the code, a number of benchmark simulations were performed with our latest version of CHIEF, using the setup of the model described in Section~\ref{sec:setup}. In addition to the setup with $N_x \times N_y \times N_z = 256 \times 256 \times 256$ grid points (denoted $256^3$ for short) used for \textit{Run 1} and \textit{Run 2}, we show, for comparison, also benchmarks for an 8-fold smaller ($128^3$) and an 8-fold larger ($512^3$) number of grid points. In all cases, a number of 60 particles per cell was used, according to Section~\ref{sec:setup}. The performance characteristics of CHIEF for quasi-2D simulations were already documented in \citet{Jain2022}.

The computational grid for the 3D simulations and benchmarks is partitioned across MPI processes using a three-dimensional domain decomposition, and assigning an equal-as-possible number of grid points and 
particles to each MPI rank (or processor core).
All calculations were performed on the HPC system \textit{Raven} \cite{MPCDF_Raven_doc} at the Max Planck Computing and Data Facility, the compute nodes of which provide two Intel Xeon Platinum 8360Y (IceLake-SP) processors with 72 cores per node, operated at the nominal frequency of 2.4 GHz (turbo mode disabled). The nodes are connected through a 100 Gb/s Nvidia Mellanox HDR100 InfiniBand network with a non-blocking fat-tree topology. 
We employed Intel 2021.6 C++ and Fortran compilers, and Intel MPI 2021.6, Intel MKL 2022.1, Blitz++ 1.0, and HYPRE 2.27 libraries. Out of several alternatives offered by HYPRE \cite{falgout2002}, the hybrid solver (BiCGSTAB with a single step of BoomerAMG preconditioning) was found to provide the best performance for our setup.

The run time per time step serves as our primary performance metric. It is calculated as the arithmetic average of the first five time steps in three different runs (run-to-run variation of run times were negligible). These time steps are performed during a very early epoch of the simulation when the spatial distribution of particles is still homogeneous and thus PIC-specific imbalances of the workload per processor core are still absent. Later in the simulations, such load imbalances are very likely responsible for an increase of the run time per time step by up to 40\%. We note that in its current version, the CHIEF code does not yet implement a particle-redistribution mechanism for re-balancing the per-processor workloads.

\begin{figure}[ht!]
	\begin{minipage}[b]{0.49\textwidth}
		\includegraphics[width=0.99\textwidth]{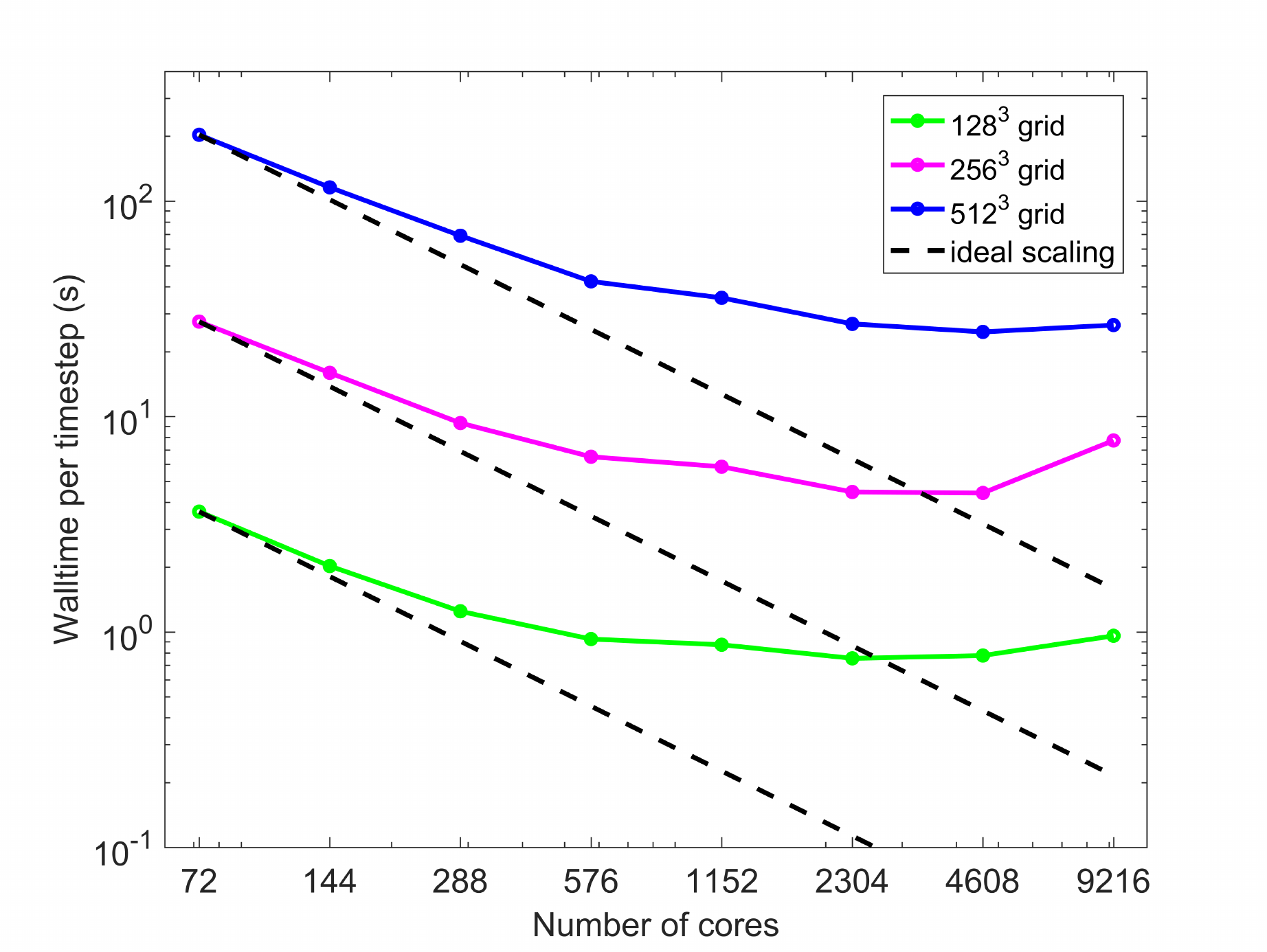}
	\end{minipage}
	\begin{minipage}[b]{0.49\textwidth}
		\includegraphics[width=0.99\textwidth]{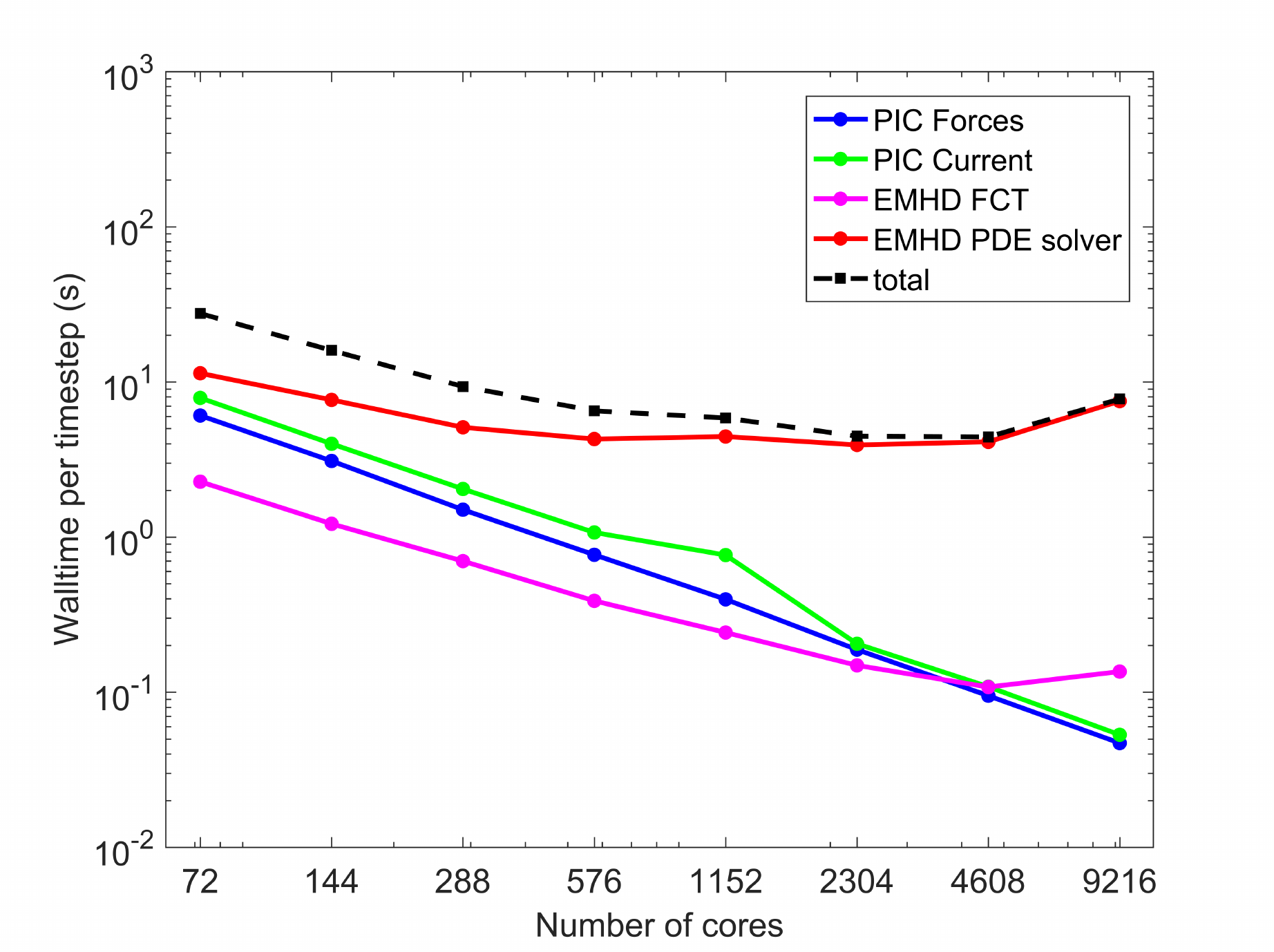}
	\end{minipage}
\caption{Strong scaling results (run time as a function of the number of processor cores) for the CHIEF code using up to 9216 processor cores (left panel), and a breakdown into its main algorithmic parts for the $256^3$-grid with 60 particles per cell (right panel). "PIC Forces": moving particles via the Boris pusher, "PIC Current": depositing the current due to particles onto the grid. "EMHD": electron-magnetohydrodynamics (Maxwell equations and electron fluid equations), "FCT": flux corrected transport (part of the EMHD algorithm that solves a generalized continuity equation), "EMHD PDE solver": HYPRE solver.
\label{fig:strongscaling}}
\end{figure}

The left panel of Fig.~\ref{fig:strongscaling} shows the run time per time step as a function of the number of processor cores (resp. MPI tasks) for the three different grids (distinguished by different colours) with up to 9216 cores (128 nodes). The black lines indicate ideal strong scaling, taking the computation time on a single node (72 cores) as the reference.
The code maintains good strong scalability up to roughly 576 cores for all the benchmarked grids, and, for the largest ($512^{3}$) grid, is able to still decrease the run time down to ca. 26 s per time step when using 9216 cores. As expected, deviations from ideal scaling becomes more prominent with smaller numbers of grid points. These benchmarks motivated our choice of running the production simulations (with a $256^{3}$-grid and 60 particles per cell, see above) shown in this paper on up to a maximum of ca.~1200 processor cores. The longest run took approximately 120 (wallclock) hours.

\begin{figure}[!ht]
	\begin{minipage}[b]{0.49\textwidth}
		\includegraphics[width=0.99\textwidth]{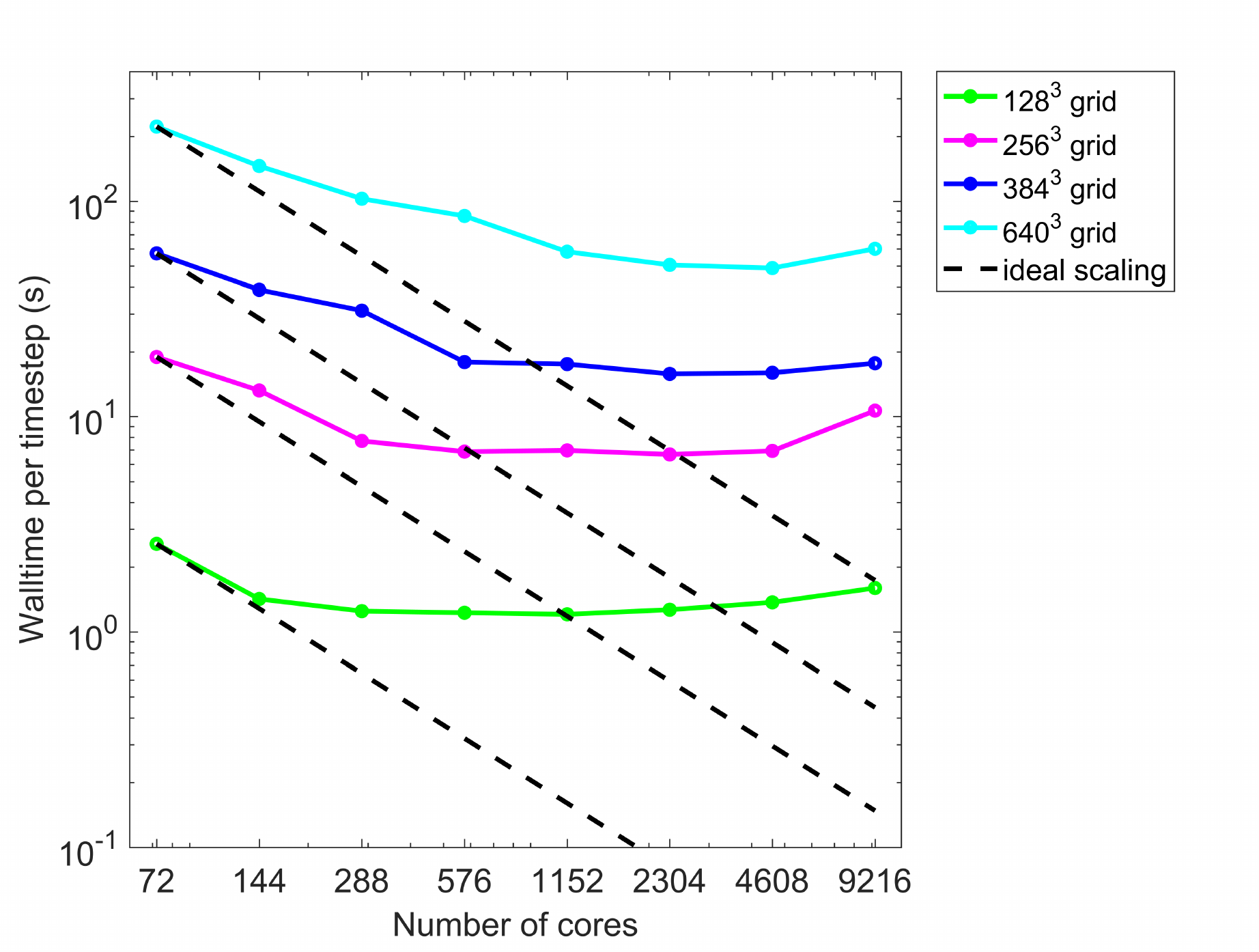}
	\end{minipage}
  \caption{Strong scaling results (run time as a function of the number of processor cores) for the HYPRE solver in the EMHD part of the CHIEF code, using up to 9216 cores (128 nodes) for different numbers of grid points.
  \label{fig:hypre_scaling}}
\end{figure}

For the $256^{3}$-grid used in this study, the right panel of Fig.~\ref{fig:strongscaling} shows the strong scaling behaviour of CHIEF with a breakdown into the main algorithmic parts (detailed in \citet{Munoz2018,Jain2022}).  
%
When running on a single compute node, the overall performance of the code is mainly limited by the available memory bandwidth, which is due to the relatively low algorithmic intensity of 0.77 (a value of at least 20 would be required to achieve peak performance on a single compute node of \textit{Raven}, according to a basic roofline model) that is achievable by such type of code. When increasing the number of nodes, while the PIC parts of the algorithm scale virtually perfectly in the studied range, the overall strong scalability gets increasingly limited by the less scalable PDE solver in the EMHD part of the code. A detailed analysis has shown that the HYPRE solver is plagued by a significant intrinsic load imbalance beyond a few hundred processor cores and is thus identified as the main scaling bottleneck for the given setup. 
This is further illustrated by Fig.~\ref{fig:hypre_scaling} which shows the run time per time step only for the HYPRE PDE solver as a function of the number of processor cores for different number of grid points, up to a maximum of $640^3$ which we could fit into the available memory of the nodes. As expected, the parallel scalability of the HYPRE solver improves (moderately) with increasing number of grid points.

Obviously, shifting the relative computational load towards the well-scaling PIC part by increasing the number of particles per cell will always lead to an improvement of the overall parallel scalability. Therefore, if CHIEF simulations with a number of particles substantially higher than used in this study are desired, the corresponding increase in the (wallclock) run time can be contained by running with a higher parallel efficiency on a larger number (beyond $10,000$) of processor cores.

\section{Energy conservation and partition \label{sec:energy}}

\begin{figure}[!ht]
                    \includegraphics[width=0.99\textwidth]{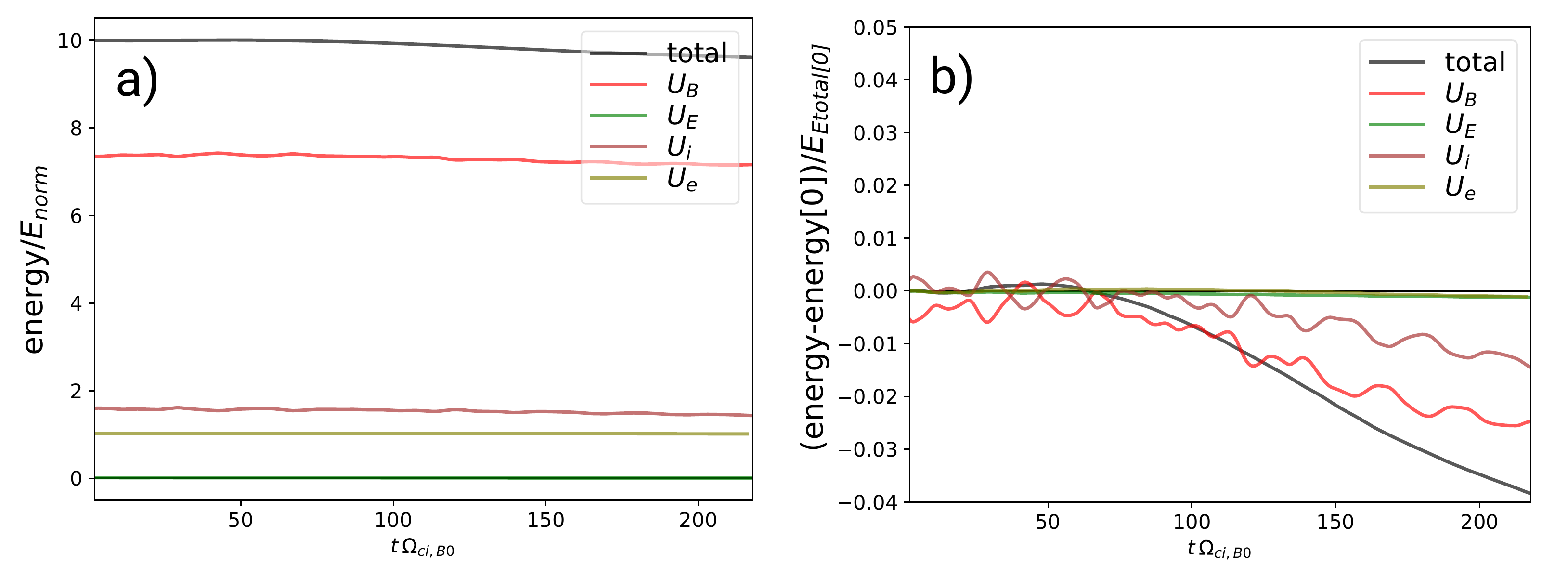}
                        \caption{Time traces of the total energy and their components for Run 1.
                        a) Energy components normalized to the initial (ion or electron) thermal energy.
                        b) Energy components normalized as $(U-U[0])/U_{total}[0]$, in order to show the variations around their initial values
                        \label{fig:energy_evolution}
                        }
                \end{figure}

For the simulations of the present manuscript, the total energy is conserved within 1\% at the time of diagnostics, while it is within 4\% of its initial value at the end of the simulations (See Fig. \ref{fig:energy_evolution}b). The reduction in the total energy is probably due to the dissipation of turbulence. 

The total energy in Fig. \ref{fig:energy_evolution} (as well as in our paper with the original description of our code \citep{Munoz2018}) is calculated as the sum of the following energy components:

\begin{equation}
    U_{total} = U_B + U_E + U_i + U_e
\end{equation}
where $U_B$ is the magnetic field energy, $U_E$ is the electric field energy, $U_i$ is the total ion kinetic energy, $U_e$ is the total electron kinetic energy. 
Fig.\ref{fig:energy_evolution}a) shows how the magnetic energy dominates the other components, obeying the low-beta plasma conditions. The total ion kinetic energy is larger than the electric kinetic energy, while the electric field energy is negligible in comparison with all the other terms. Note in Fig.\ref{fig:energy_evolution}b) how the ion kinetic energy is transformed into magnetic field energy and vice versa.

In the following we discuss the calculation of each one of the individual energy components. The electromagnetic field energies are defined as:
\begin{align}
    U_B&= \int \frac{1}{2\mu_0}B^2\;d^3\vec{x} = \sum_{i=1}^{N_x} \sum_{j=1}^{N_y} \sum_{k=1}^{N_z} \Delta x^3 \frac{1}{2\mu_0}B_{i,j,k}^2\\
    U_E&= \int \frac{\epsilon_0}{2}E^2\;d^3\vec{x}  = \sum_{i=1}^{N_x} \sum_{j=1}^{N_y} \sum_{k=1}^{N_z} \Delta x^3  \frac{\epsilon_0}{2}E_{i,j,k}^2
\end{align}
where the terms after the first equal sign represent the continuum definition and the terms after the second equal sign represent the discrete approximation used for the calculations. The indices $i,j,k$ represent the grid points, and $N_x, N_y, N_z$ the number of grid points along each direction in the simulation domain.

In order to determine the kinetic energies $U_i$ or $U_e$, we first calculate the ion kinetic energy density per cell $u_i$ (the electron energy is obtained by replacing the subscripts $i$ by $e$) defined as:
\begin{align}\label{eq:ui_per_cell}
u_i(\vec{x}) &= \frac{1}{\Delta x^3}\sum_{p=1}^{N_p} \frac{1}{2}M_i \vec{v}_p^2 =  \frac{1}{2}m_in_i(\vec{x})\frac{1}{N_p}\sum_{p=1}^{N_p}  v_p^2\\
&=\frac{1}{2}m_i\int \vec{v}^2\;f_i(\vec{x},\vec{v})\;d^3\vec{v}\\
&=\frac{1}{2}m_in_i(\vec{x})\left[ \vec{V_i}(\vec{x})^2 +  \frac{1}{n_i} \int (\vec{v} - \vec{V}_i(\vec{x}))^2 \;f_i(\vec{x},\vec{v})\;d^3\vec{v}  \right]\\
&=u_{k,i} + u_{th,i}\label{eq:last}
\end{align}

where $p$ is the particle index running over all the macro-particles in one cell, $N_p$ the number of (ion) macro-particles in one cell,
and $M_i$ is the mass of an ion macro-particle, which is related to the ion mass via $M_i=(n_i\Delta x^3/N_p) m_i $, where $m_i$ is the mass of a proton (ion), $n_i$ is the physical density, and the factor in front of $m_i$ is usually called macro-factor, representing the ratio of physical to numerical particles.
The equality from the first to the second row represents a transition from the discrete to the continuum representation. The thermal kinetic energy density and bulk flow kinetic energy densities are then defined from Eq.~\eqref{eq:last} respectively as:
\begin{align}
u_{k,i}(\vec{x}) & = \frac{1}{2}m_in_i(\vec{x})\left[ V_{i,x}(\vec{x})^2 + V_{i,y}(\vec{x})^2 + V_{i,z}(\vec{x})^2 \right] \label{eq:uki_percell}\\
u_{th,i}(\vec{x}) & = \frac{1}{2}n_i(\vec{x})k_B\left[ T_{i,x}(\vec{x}) + T_{i,y}(\vec{x}) + T_{i,z}(\vec{x}) \right] \label{eq:uthi_percell}
\end{align}
where the bulk flow energies are defined as 
\begin{align}\label{eq:vi}
\vec{V}_i(\vec{x}) =  \frac{1}{n_i}\int \vec{v}\;f_i\;d^3\vec{v}= \frac{1}{N_p}\sum_{p=1}^{N_p} \vec{v}_p  
\end{align}
and the temperature components as 
\begin{align}\label{eq:ti}
k_BT_{i,l}(\vec{x}) =  \frac{m_i}{n_i} \int (v_l - V_{i,l})^2\;f_i\;d^3\vec{v}=\frac{m_i }{N_p}\sum_{p=1}^{N_p} (v_{p,l} - V_{i,l})^2
\end{align}
where the index $l$ represents each spatial direction. Note that the terms after the second equal sign in Eq.~\eqref{eq:vi} and Eq.~\eqref{eq:uthi} represent their discrete version in each cell, but the code calculates those quantities with a weighted average on each cell via a second order shape function.

The total kinetic energy is then the integral of the total kinetic energy density Eq.~\eqref{eq:ui_per_cell} over all the simulation domain:
\begin{align}\label{eq:ion_total_kinetic_energy}
    U_i &=\int u_i(\vec{x}) d^3\vec{x} = \sum_{p=1}^{N} \frac{1}{2}M_i \vec{v}_p^2 
    =\int \left[u_{k,i}(\vec{x}) + u_{th,i}(\vec{x}) \right]d^3\vec{x} 
\end{align}
where the index $p$ now runs over all the $N$ particles in the simulation domain. Its components in the discrete version are obtained by integrating the densities in Eqs.~\eqref{eq:uki_percell} and \eqref{eq:uthi_percell} over all the simulation domain:
\begin{align} \label{eq:uki}
U_{k,i} & = \int u_{k,i}(\vec{x}) d^3\vec{x}
=\frac{1}{2}m_i \Delta x^3 \sum_{i=1}^{N_x} \sum_{j=1}^{N_y} \sum_{k=1}^{N_z}  n_{i,j,k}\left[V_{ix,i,j,k}^2 +V_{iy,i,j,k}^2 +V_{iz,i,j,k}^2 \right] \\
U_{th,i} & = \int u_{th,i}(\vec{x}) d^3\vec{x}
=\frac{1}{2} \Delta x^3\sum_{i=1}^{N_x} \sum_{j=1}^{N_y} \sum_{k=1}^{N_z}  n_{i,j,k}k_B\left[T_{ix,i,j,k} + T_{iy,i,j,k} +T_{iz,i,j,k} \right] \label{eq:uthi}
\end{align}

For our calculations, we used the particle definition Eq.~\ref{eq:ion_total_kinetic_energy} for the total ion kinetic energy shown in Fig.~\ref{fig:energy_evolution}.
This is approximately equal to the field definition $U_{k,i} + U_{th,i}$.
For the fluid electrons, we have to use the field definitions of its components bulk flow energy as per Eq.~\eqref{eq:uki} with the indices for ions changed to electrons:
\begin{align} \label{eq:uke}
U_{k,e} & =\frac{1}{2}m_e \Delta x^3 \sum_{i=1}^{N_x} \sum_{j=1}^{N_y} \sum_{k=1}^{N_z}  n_{e,j,k}\left[V_{ex,i,j,k}^2 +V_{ey,i,j,k}^2 +V_{ez,i,j,k}^2 \right]
\end{align}
where $n_{e,j,k}=n_{i,j,k}$ and thermal energy as in Eq.~\eqref{eq:uthi} for electrons. Since the electron temperature is constant and isotropic, Eq.~\eqref{eq:uthi}  simplifies to \begin{equation}
U_{th,e}  =\frac{3}{2} \Delta x^3\sum_{i=1}^{N_x} \sum_{j=1}^{N_y} \sum_{k=1}^{N_z}  n_{i,j,k}k_B T_{e,i,j,k} \label{eq:uthe}
\end{equation}

Note that in the limit of massless electrons, the bulk flow kinetic energy of electrons Eq.~\eqref{eq:uke} becomes zero, but the thermal energy Eq.~\eqref{eq:uthe} does not change.

\bibliography{3DparallelCHIEF}

\begin{thebibliography}{65}%
\makeatletter
\providecommand \@ifxundefined [1]{%
 \@ifx{#1\undefined}
}%
\providecommand \@ifnum [1]{%
 \ifnum #1\expandafter \@firstoftwo
 \else \expandafter \@secondoftwo
 \fi
}%
\providecommand \@ifx [1]{%
 \ifx #1\expandafter \@firstoftwo
 \else \expandafter \@secondoftwo
 \fi
}%
\providecommand \natexlab [1]{#1}%
\providecommand \enquote  [1]{``#1''}%
\providecommand \bibnamefont  [1]{#1}%
\providecommand \bibfnamefont [1]{#1}%
\providecommand \citenamefont [1]{#1}%
\providecommand \href@noop [0]{\@secondoftwo}%
\providecommand \href [0]{\begingroup \@sanitize@url \@href}%
\providecommand \@href[1]{\@@startlink{#1}\@@href}%
\providecommand \@@href[1]{\endgroup#1\@@endlink}%
\providecommand \@sanitize@url [0]{\catcode `\\12\catcode `\$12\catcode
  `\&12\catcode `\#12\catcode `\^12\catcode `\_12\catcode `\%12\relax}%
\providecommand \@@startlink[1]{}%
\providecommand \@@endlink[0]{}%
\providecommand \url  [0]{\begingroup\@sanitize@url \@url }%
\providecommand \@url [1]{\endgroup\@href {#1}{\urlprefix }}%
\providecommand \urlprefix  [0]{URL }%
\providecommand \Eprint [0]{\href }%
\providecommand \doibase [0]{http://dx.doi.org/}%
\providecommand \selectlanguage [0]{\@gobble}%
\providecommand \bibinfo  [0]{\@secondoftwo}%
\providecommand \bibfield  [0]{\@secondoftwo}%
\providecommand \translation [1]{[#1]}%
\providecommand \BibitemOpen [0]{}%
\providecommand \bibitemStop [0]{}%
\providecommand \bibitemNoStop [0]{.\EOS\space}%
\providecommand \EOS [0]{\spacefactor3000\relax}%
\providecommand \BibitemShut  [1]{\csname bibitem#1\endcsname}%
\let\auto@bib@innerbib\@empty
\bibitem [{\citenamefont {Bruno}\ and\ \citenamefont
  {Carbone}(2013)}]{Bruno2013}%
  \BibitemOpen
  \bibfield  {author} {\bibinfo {author} {\bibfnamefont {R.}~\bibnamefont
  {Bruno}}\ and\ \bibinfo {author} {\bibfnamefont {V.}~\bibnamefont
  {Carbone}},\ }\href {\doibase 10.12942/lrsp-2013-2} {\bibfield  {journal}
  {\bibinfo  {journal} {Living Rev. Sol. Phys.}\ }\textbf {\bibinfo {volume}
  {10}},\ \bibinfo {pages} {2} (\bibinfo {year} {2013})}\BibitemShut {NoStop}%
\bibitem [{\citenamefont {Howes}(2015{\natexlab{a}})}]{Howes2015a}%
  \BibitemOpen
  \bibfield  {author} {\bibinfo {author} {\bibfnamefont {G.~G.}\ \bibnamefont
  {Howes}},\ }in\ \href {\doibase 10.1007/978-3-662-44625-6_6} {\emph {\bibinfo
  {booktitle} {Magnetic Fields in Diffuse Media}}},\ \bibinfo {series}
  {Astrophysics and Space Science Library}, Vol.\ \bibinfo {volume} {407},\
  \bibinfo {editor} {edited by\ \bibinfo {editor} {\bibfnamefont
  {A.}~\bibnamefont {Lazarian}}, \bibinfo {editor} {\bibfnamefont {E.~M.}\
  \bibnamefont {{de Gouveia Dal Pino}}}, \ and\ \bibinfo {editor}
  {\bibfnamefont {C.}~\bibnamefont {Melioli}}}\ (\bibinfo  {publisher}
  {Springer Berlin Heidelberg},\ \bibinfo {address} {Berlin, Heidelberg},\
  \bibinfo {year} {2015})\ pp.\ \bibinfo {pages} {123--152}\BibitemShut
  {NoStop}%
\bibitem [{\citenamefont {Chen}(2016)}]{Chen2017y}%
  \BibitemOpen
  \bibfield  {author} {\bibinfo {author} {\bibfnamefont {C.~H.~K.}\
  \bibnamefont {Chen}},\ }\href {\doibase 10.1017/S0022377816001124} {\bibfield
   {journal} {\bibinfo  {journal} {J. Plasma Phys.}\ }\textbf {\bibinfo
  {volume} {82}},\ \bibinfo {pages} {535820602} (\bibinfo {year}
  {2016})}\BibitemShut {NoStop}%
\bibitem [{\citenamefont {Verscharen}\ \emph {et~al.}(2019)\citenamefont
  {Verscharen}, \citenamefont {Klein},\ and\ \citenamefont
  {Maruca}}]{Verscharen2019c}%
  \BibitemOpen
  \bibfield  {author} {\bibinfo {author} {\bibfnamefont {D.}~\bibnamefont
  {Verscharen}}, \bibinfo {author} {\bibfnamefont {K.~G.}\ \bibnamefont
  {Klein}}, \ and\ \bibinfo {author} {\bibfnamefont {B.~A.}\ \bibnamefont
  {Maruca}},\ }\href {\doibase 10.1007/s41116-019-0021-0} {\bibfield  {journal}
  {\bibinfo  {journal} {Living Rev. Sol. Phys.}\ }\textbf {\bibinfo {volume}
  {16}},\ \bibinfo {pages} {5} (\bibinfo {year} {2019})}\BibitemShut {NoStop}%
\bibitem [{\citenamefont {Matthaeus}\ and\ \citenamefont
  {Velli}(2011)}]{Matthaeus2011}%
  \BibitemOpen
  \bibfield  {author} {\bibinfo {author} {\bibfnamefont {W.~H.}\ \bibnamefont
  {Matthaeus}}\ and\ \bibinfo {author} {\bibfnamefont {M.}~\bibnamefont
  {Velli}},\ }\href {\doibase 10.1007/s11214-011-9793-9} {\bibfield  {journal}
  {\bibinfo  {journal} {Space Sci. Rev.}\ }\textbf {\bibinfo {volume} {160}},\
  \bibinfo {pages} {145} (\bibinfo {year} {2011})}\BibitemShut {NoStop}%
\bibitem [{\citenamefont {Karimabadi}\ \emph {et~al.}(2013)\citenamefont
  {Karimabadi}, \citenamefont {Roytershteyn}, \citenamefont {Daughton},\ and\
  \citenamefont {Liu}}]{Karimabadi2013}%
  \BibitemOpen
  \bibfield  {author} {\bibinfo {author} {\bibfnamefont {H.}~\bibnamefont
  {Karimabadi}}, \bibinfo {author} {\bibfnamefont {V.}~\bibnamefont
  {Roytershteyn}}, \bibinfo {author} {\bibfnamefont {W.}~\bibnamefont
  {Daughton}}, \ and\ \bibinfo {author} {\bibfnamefont {Y.-H.}\ \bibnamefont
  {Liu}},\ }\href {\doibase 10.1007/s11214-013-0021-7} {\bibfield  {journal}
  {\bibinfo  {journal} {Space Sci. Rev.}\ }\textbf {\bibinfo {volume} {178}},\
  \bibinfo {pages} {307} (\bibinfo {year} {2013})}\BibitemShut {NoStop}%
\bibitem [{\citenamefont {Treumann}\ and\ \citenamefont
  {Baumjohann}(2013)}]{Treumann2013b}%
  \BibitemOpen
  \bibfield  {author} {\bibinfo {author} {\bibfnamefont {R.~A.}\ \bibnamefont
  {Treumann}}\ and\ \bibinfo {author} {\bibfnamefont {W.}~\bibnamefont
  {Baumjohann}},\ }\href {\doibase 10.3389/fphy.2013.00031} {\bibfield
  {journal} {\bibinfo  {journal} {Front. Phys.}\ }\textbf {\bibinfo {volume}
  {1}},\ \bibinfo {pages} {31} (\bibinfo {year} {2013})}\BibitemShut {NoStop}%
\bibitem [{\citenamefont {Matthaeus}\ and\ \citenamefont
  {Lamkin}(1986)}]{Matthaeus1986}%
  \BibitemOpen
  \bibfield  {author} {\bibinfo {author} {\bibfnamefont {W.~H.}\ \bibnamefont
  {Matthaeus}}\ and\ \bibinfo {author} {\bibfnamefont {S.~L.}\ \bibnamefont
  {Lamkin}},\ }\href {\doibase 10.1063/1.866004} {\bibfield  {journal}
  {\bibinfo  {journal} {Phys. Fluids}\ }\textbf {\bibinfo {volume} {29}},\
  \bibinfo {pages} {2513} (\bibinfo {year} {1986})}\BibitemShut {NoStop}%
\bibitem [{\citenamefont {Servidio}\ \emph {et~al.}(2010)\citenamefont
  {Servidio}, \citenamefont {Matthaeus}, \citenamefont {Shay}, \citenamefont
  {Dmitruk}, \citenamefont {Cassak},\ and\ \citenamefont {Wan}}]{Servidio2010}%
  \BibitemOpen
  \bibfield  {author} {\bibinfo {author} {\bibfnamefont {S.}~\bibnamefont
  {Servidio}}, \bibinfo {author} {\bibfnamefont {W.~H.}\ \bibnamefont
  {Matthaeus}}, \bibinfo {author} {\bibfnamefont {M.~A.}\ \bibnamefont {Shay}},
  \bibinfo {author} {\bibfnamefont {P.}~\bibnamefont {Dmitruk}}, \bibinfo
  {author} {\bibfnamefont {P.~A.}\ \bibnamefont {Cassak}}, \ and\ \bibinfo
  {author} {\bibfnamefont {M.}~\bibnamefont {Wan}},\ }\href {\doibase
  10.1063/1.3368798} {\bibfield  {journal} {\bibinfo  {journal} {Phys.
  Plasmas}\ }\textbf {\bibinfo {volume} {17}},\ \bibinfo {pages} {032315}
  (\bibinfo {year} {2010})}\BibitemShut {NoStop}%
\bibitem [{\citenamefont {Franci}\ \emph {et~al.}(2017)\citenamefont {Franci},
  \citenamefont {Cerri}, \citenamefont {Califano}, \citenamefont {Landi},
  \citenamefont {Papini}, \citenamefont {Verdini}, \citenamefont {Matteini},
  \citenamefont {Jenko},\ and\ \citenamefont {Hellinger}}]{Franci2017}%
  \BibitemOpen
  \bibfield  {author} {\bibinfo {author} {\bibfnamefont {L.}~\bibnamefont
  {Franci}}, \bibinfo {author} {\bibfnamefont {S.~S.}\ \bibnamefont {Cerri}},
  \bibinfo {author} {\bibfnamefont {F.}~\bibnamefont {Califano}}, \bibinfo
  {author} {\bibfnamefont {S.}~\bibnamefont {Landi}}, \bibinfo {author}
  {\bibfnamefont {E.}~\bibnamefont {Papini}}, \bibinfo {author} {\bibfnamefont
  {A.}~\bibnamefont {Verdini}}, \bibinfo {author} {\bibfnamefont
  {L.}~\bibnamefont {Matteini}}, \bibinfo {author} {\bibfnamefont
  {F.}~\bibnamefont {Jenko}}, \ and\ \bibinfo {author} {\bibfnamefont
  {P.}~\bibnamefont {Hellinger}},\ }\href {\doibase 10.3847/2041-8213/aa93fb}
  {\bibfield  {journal} {\bibinfo  {journal} {Astrophys. J.}\ }\textbf
  {\bibinfo {volume} {850}},\ \bibinfo {pages} {L16} (\bibinfo {year}
  {2017})}\BibitemShut {NoStop}%
\bibitem [{\citenamefont {Cerri}\ and\ \citenamefont
  {Califano}(2017)}]{Cerri2017}%
  \BibitemOpen
  \bibfield  {author} {\bibinfo {author} {\bibfnamefont {S.~S.}\ \bibnamefont
  {Cerri}}\ and\ \bibinfo {author} {\bibfnamefont {F.}~\bibnamefont
  {Califano}},\ }\href {\doibase 10.1088/1367-2630/aa5c4a} {\bibfield
  {journal} {\bibinfo  {journal} {New J. Phys.}\ }\textbf {\bibinfo {volume}
  {19}},\ \bibinfo {pages} {025007} (\bibinfo {year} {2017})}\BibitemShut
  {NoStop}%
\bibitem [{\citenamefont {Haggerty}\ \emph {et~al.}(2017)\citenamefont
  {Haggerty}, \citenamefont {Parashar}, \citenamefont {Matthaeus},
  \citenamefont {Shay}, \citenamefont {Yang}, \citenamefont {Wan},
  \citenamefont {Wu},\ and\ \citenamefont {Servidio}}]{Haggerty2017a}%
  \BibitemOpen
  \bibfield  {author} {\bibinfo {author} {\bibfnamefont {C.~C.}\ \bibnamefont
  {Haggerty}}, \bibinfo {author} {\bibfnamefont {T.~N.}\ \bibnamefont
  {Parashar}}, \bibinfo {author} {\bibfnamefont {W.~H.}\ \bibnamefont
  {Matthaeus}}, \bibinfo {author} {\bibfnamefont {M.~A.}\ \bibnamefont {Shay}},
  \bibinfo {author} {\bibfnamefont {Y.}~\bibnamefont {Yang}}, \bibinfo {author}
  {\bibfnamefont {M.}~\bibnamefont {Wan}}, \bibinfo {author} {\bibfnamefont
  {P.}~\bibnamefont {Wu}}, \ and\ \bibinfo {author} {\bibfnamefont
  {S.}~\bibnamefont {Servidio}},\ }\href {\doibase 10.1063/1.5001722}
  {\bibfield  {journal} {\bibinfo  {journal} {Phys. Plasmas}\ }\textbf
  {\bibinfo {volume} {24}},\ \bibinfo {pages} {102308} (\bibinfo {year}
  {2017})}\BibitemShut {NoStop}%
\bibitem [{\citenamefont {Jain}\ \emph {et~al.}(2021)\citenamefont {Jain},
  \citenamefont {Büchner}, \citenamefont {Comi{\c{s}}el},\ and\ \citenamefont
  {Motschmann}}]{Jain2021}%
  \BibitemOpen
  \bibfield  {author} {\bibinfo {author} {\bibfnamefont {N.}~\bibnamefont
  {Jain}}, \bibinfo {author} {\bibfnamefont {J.}~\bibnamefont {Büchner}},
  \bibinfo {author} {\bibfnamefont {H.}~\bibnamefont {Comi{\c{s}}el}}, \ and\
  \bibinfo {author} {\bibfnamefont {U.}~\bibnamefont {Motschmann}},\ }\href
  {\doibase 10.3847/1538-4357/ac106c} {\bibfield  {journal} {\bibinfo
  {journal} {The Astrophysical Journal}\ }\textbf {\bibinfo {volume} {919}},\
  \bibinfo {pages} {103} (\bibinfo {year} {2021})}\BibitemShut {NoStop}%
\bibitem [{\citenamefont {Yordanova}\ \emph {et~al.}(2016)\citenamefont
  {Yordanova}, \citenamefont {V{\"{o}}r{\"{o}}s}, \citenamefont {Varsani},
  \citenamefont {Graham}, \citenamefont {Norgren}, \citenamefont
  {Khotyaintsev}, \citenamefont {Vaivads}, \citenamefont {Eriksson},
  \citenamefont {Nakamura}, \citenamefont {Lindqvist}, \citenamefont
  {Marklund}, \citenamefont {Ergun}, \citenamefont {Magnes}, \citenamefont
  {Baumjohann}, \citenamefont {Fischer}, \citenamefont {Plaschke},
  \citenamefont {Narita}, \citenamefont {Russell}, \citenamefont {Strangeway},
  \citenamefont {{Le Contel}}, \citenamefont {Pollock}, \citenamefont
  {Torbert}, \citenamefont {Giles}, \citenamefont {Burch}, \citenamefont
  {Avanov}, \citenamefont {Dorelli}, \citenamefont {Gershman}, \citenamefont
  {Paterson}, \citenamefont {Lavraud},\ and\ \citenamefont
  {Saito}}]{Yordanova2016}%
  \BibitemOpen
  \bibfield  {author} {\bibinfo {author} {\bibfnamefont {E.}~\bibnamefont
  {Yordanova}}, \bibinfo {author} {\bibfnamefont {Z.}~\bibnamefont
  {V{\"{o}}r{\"{o}}s}}, \bibinfo {author} {\bibfnamefont {A.}~\bibnamefont
  {Varsani}}, \bibinfo {author} {\bibfnamefont {D.~B.}\ \bibnamefont {Graham}},
  \bibinfo {author} {\bibfnamefont {C.}~\bibnamefont {Norgren}}, \bibinfo
  {author} {\bibfnamefont {Y.~V.}\ \bibnamefont {Khotyaintsev}}, \bibinfo
  {author} {\bibfnamefont {A.}~\bibnamefont {Vaivads}}, \bibinfo {author}
  {\bibfnamefont {E.}~\bibnamefont {Eriksson}}, \bibinfo {author}
  {\bibfnamefont {R.}~\bibnamefont {Nakamura}}, \bibinfo {author}
  {\bibfnamefont {P.-A.}\ \bibnamefont {Lindqvist}}, \bibinfo {author}
  {\bibfnamefont {G.}~\bibnamefont {Marklund}}, \bibinfo {author}
  {\bibfnamefont {R.~E.}\ \bibnamefont {Ergun}}, \bibinfo {author}
  {\bibfnamefont {W.}~\bibnamefont {Magnes}}, \bibinfo {author} {\bibfnamefont
  {W.}~\bibnamefont {Baumjohann}}, \bibinfo {author} {\bibfnamefont
  {D.}~\bibnamefont {Fischer}}, \bibinfo {author} {\bibfnamefont
  {F.}~\bibnamefont {Plaschke}}, \bibinfo {author} {\bibfnamefont
  {Y.}~\bibnamefont {Narita}}, \bibinfo {author} {\bibfnamefont {C.~T.}\
  \bibnamefont {Russell}}, \bibinfo {author} {\bibfnamefont {R.~J.}\
  \bibnamefont {Strangeway}}, \bibinfo {author} {\bibfnamefont
  {O.}~\bibnamefont {{Le Contel}}}, \bibinfo {author} {\bibfnamefont
  {C.}~\bibnamefont {Pollock}}, \bibinfo {author} {\bibfnamefont {R.~B.}\
  \bibnamefont {Torbert}}, \bibinfo {author} {\bibfnamefont {B.~J.}\
  \bibnamefont {Giles}}, \bibinfo {author} {\bibfnamefont {J.~L.}\ \bibnamefont
  {Burch}}, \bibinfo {author} {\bibfnamefont {L.~A.}\ \bibnamefont {Avanov}},
  \bibinfo {author} {\bibfnamefont {J.~C.}\ \bibnamefont {Dorelli}}, \bibinfo
  {author} {\bibfnamefont {D.~J.}\ \bibnamefont {Gershman}}, \bibinfo {author}
  {\bibfnamefont {W.~R.}\ \bibnamefont {Paterson}}, \bibinfo {author}
  {\bibfnamefont {B.}~\bibnamefont {Lavraud}}, \ and\ \bibinfo {author}
  {\bibfnamefont {Y.}~\bibnamefont {Saito}},\ }\href {\doibase
  10.1002/2016GL069191} {\bibfield  {journal} {\bibinfo  {journal} {Geophys.
  Res. Lett.}\ }\textbf {\bibinfo {volume} {43}},\ \bibinfo {pages} {5969}
  (\bibinfo {year} {2016})}\BibitemShut {NoStop}%
\bibitem [{\citenamefont {V{\"{o}}r{\"{o}}s}\ \emph {et~al.}(2017)\citenamefont
  {V{\"{o}}r{\"{o}}s}, \citenamefont {Yordanova}, \citenamefont {Varsani},
  \citenamefont {Genestreti}, \citenamefont {Khotyaintsev}, \citenamefont {Li},
  \citenamefont {Graham}, \citenamefont {Norgren}, \citenamefont {Nakamura},
  \citenamefont {Narita}, \citenamefont {Plaschke}, \citenamefont {Magnes},
  \citenamefont {Baumjohann}, \citenamefont {Fischer}, \citenamefont {Vaivads},
  \citenamefont {Eriksson}, \citenamefont {Lindqvist}, \citenamefont
  {Marklund}, \citenamefont {Ergun}, \citenamefont {Leitner}, \citenamefont
  {Leubner}, \citenamefont {Strangeway}, \citenamefont {{Le Contel}},
  \citenamefont {Pollock}, \citenamefont {Giles}, \citenamefont {Torbert},
  \citenamefont {Burch}, \citenamefont {Avanov}, \citenamefont {Dorelli},
  \citenamefont {Gershman}, \citenamefont {Paterson}, \citenamefont {Lavraud},\
  and\ \citenamefont {Saito}}]{Voros2017}%
  \BibitemOpen
  \bibfield  {author} {\bibinfo {author} {\bibfnamefont {Z.}~\bibnamefont
  {V{\"{o}}r{\"{o}}s}}, \bibinfo {author} {\bibfnamefont {E.}~\bibnamefont
  {Yordanova}}, \bibinfo {author} {\bibfnamefont {A.}~\bibnamefont {Varsani}},
  \bibinfo {author} {\bibfnamefont {K.~J.}\ \bibnamefont {Genestreti}},
  \bibinfo {author} {\bibfnamefont {Y.~V.}\ \bibnamefont {Khotyaintsev}},
  \bibinfo {author} {\bibfnamefont {W.}~\bibnamefont {Li}}, \bibinfo {author}
  {\bibfnamefont {D.~B.}\ \bibnamefont {Graham}}, \bibinfo {author}
  {\bibfnamefont {C.}~\bibnamefont {Norgren}}, \bibinfo {author} {\bibfnamefont
  {R.}~\bibnamefont {Nakamura}}, \bibinfo {author} {\bibfnamefont
  {Y.}~\bibnamefont {Narita}}, \bibinfo {author} {\bibfnamefont
  {F.}~\bibnamefont {Plaschke}}, \bibinfo {author} {\bibfnamefont
  {W.}~\bibnamefont {Magnes}}, \bibinfo {author} {\bibfnamefont
  {W.}~\bibnamefont {Baumjohann}}, \bibinfo {author} {\bibfnamefont
  {D.}~\bibnamefont {Fischer}}, \bibinfo {author} {\bibfnamefont
  {A.}~\bibnamefont {Vaivads}}, \bibinfo {author} {\bibfnamefont
  {E.}~\bibnamefont {Eriksson}}, \bibinfo {author} {\bibfnamefont {P.-A.}\
  \bibnamefont {Lindqvist}}, \bibinfo {author} {\bibfnamefont {G.}~\bibnamefont
  {Marklund}}, \bibinfo {author} {\bibfnamefont {R.~E.}\ \bibnamefont {Ergun}},
  \bibinfo {author} {\bibfnamefont {M.}~\bibnamefont {Leitner}}, \bibinfo
  {author} {\bibfnamefont {M.~P.}\ \bibnamefont {Leubner}}, \bibinfo {author}
  {\bibfnamefont {R.~J.}\ \bibnamefont {Strangeway}}, \bibinfo {author}
  {\bibfnamefont {O.}~\bibnamefont {{Le Contel}}}, \bibinfo {author}
  {\bibfnamefont {C.}~\bibnamefont {Pollock}}, \bibinfo {author} {\bibfnamefont
  {B.~J.}\ \bibnamefont {Giles}}, \bibinfo {author} {\bibfnamefont {R.~B.}\
  \bibnamefont {Torbert}}, \bibinfo {author} {\bibfnamefont {J.~L.}\
  \bibnamefont {Burch}}, \bibinfo {author} {\bibfnamefont {L.~A.}\ \bibnamefont
  {Avanov}}, \bibinfo {author} {\bibfnamefont {J.~C.}\ \bibnamefont {Dorelli}},
  \bibinfo {author} {\bibfnamefont {D.~J.}\ \bibnamefont {Gershman}}, \bibinfo
  {author} {\bibfnamefont {W.~R.}\ \bibnamefont {Paterson}}, \bibinfo {author}
  {\bibfnamefont {B.}~\bibnamefont {Lavraud}}, \ and\ \bibinfo {author}
  {\bibfnamefont {Y.}~\bibnamefont {Saito}},\ }\href {\doibase
  10.1002/2017JA024535} {\bibfield  {journal} {\bibinfo  {journal} {J. Geophys.
  Res. Sp. Phys.}\ }\textbf {\bibinfo {volume} {122}},\ \bibinfo {pages}
  {11,442} (\bibinfo {year} {2017})}\BibitemShut {NoStop}%
\bibitem [{\citenamefont {Khabarova}\ \emph {et~al.}(2015)\citenamefont
  {Khabarova}, \citenamefont {Zank}, \citenamefont {Li}, \citenamefont
  {le~Roux}, \citenamefont {Webb}, \citenamefont {Dosch},\ and\ \citenamefont
  {Malandraki}}]{Khabarova2015a}%
  \BibitemOpen
  \bibfield  {author} {\bibinfo {author} {\bibfnamefont {O.}~\bibnamefont
  {Khabarova}}, \bibinfo {author} {\bibfnamefont {G.~P.}\ \bibnamefont {Zank}},
  \bibinfo {author} {\bibfnamefont {G.}~\bibnamefont {Li}}, \bibinfo {author}
  {\bibfnamefont {J.~A.}\ \bibnamefont {le~Roux}}, \bibinfo {author}
  {\bibfnamefont {G.~M.}\ \bibnamefont {Webb}}, \bibinfo {author}
  {\bibfnamefont {A.}~\bibnamefont {Dosch}}, \ and\ \bibinfo {author}
  {\bibfnamefont {O.~E.}\ \bibnamefont {Malandraki}},\ }\href {\doibase
  10.1088/0004-637X/808/2/181} {\bibfield  {journal} {\bibinfo  {journal}
  {Astrophys. J.}\ }\textbf {\bibinfo {volume} {808}},\ \bibinfo {pages} {181}
  (\bibinfo {year} {2015})}\BibitemShut {NoStop}%
\bibitem [{\citenamefont {Winske}\ \emph {et~al.}(2023)\citenamefont {Winske},
  \citenamefont {Karimabadi}, \citenamefont {Le}, \citenamefont {Omidi},
  \citenamefont {Roytershteyn},\ and\ \citenamefont {Stanier}}]{Winske2023}%
  \BibitemOpen
  \bibfield  {author} {\bibinfo {author} {\bibfnamefont {D.}~\bibnamefont
  {Winske}}, \bibinfo {author} {\bibfnamefont {H.}~\bibnamefont {Karimabadi}},
  \bibinfo {author} {\bibfnamefont {A.~Y.}\ \bibnamefont {Le}}, \bibinfo
  {author} {\bibfnamefont {N.~N.}\ \bibnamefont {Omidi}}, \bibinfo {author}
  {\bibfnamefont {V.}~\bibnamefont {Roytershteyn}}, \ and\ \bibinfo {author}
  {\bibfnamefont {A.~J.}\ \bibnamefont {Stanier}},\ }\enquote {\bibinfo {title}
  {Hybrid-kinetic approach: Massless electrons},}\ in\ \href {\doibase
  10.1007/978-3-031-11870-8_3} {\emph {\bibinfo {booktitle} {Space and
  Astrophysical Plasma Simulation}}},\ \bibinfo {editor} {edited by\ \bibinfo
  {editor} {\bibfnamefont {J.}~\bibnamefont {B{\"u}chner}}}\ (\bibinfo
  {publisher} {Springer International Publishing},\ \bibinfo {year} {2023})\
  pp.\ \bibinfo {pages} {63--91}\BibitemShut {NoStop}%
\bibitem [{\citenamefont {Leamon}\ \emph {et~al.}(1998)\citenamefont {Leamon},
  \citenamefont {Smith}, \citenamefont {Ness}, \citenamefont {Matthaeus},\ and\
  \citenamefont {Wong}}]{Leamon1998}%
  \BibitemOpen
  \bibfield  {author} {\bibinfo {author} {\bibfnamefont {R.~J.}\ \bibnamefont
  {Leamon}}, \bibinfo {author} {\bibfnamefont {C.~W.}\ \bibnamefont {Smith}},
  \bibinfo {author} {\bibfnamefont {N.~F.}\ \bibnamefont {Ness}}, \bibinfo
  {author} {\bibfnamefont {W.~H.}\ \bibnamefont {Matthaeus}}, \ and\ \bibinfo
  {author} {\bibfnamefont {H.~K.}\ \bibnamefont {Wong}},\ }\href {\doibase
  10.1029/97JA03394} {\bibfield  {journal} {\bibinfo  {journal} {J. Geophys.
  Res. Sp. Phys.}\ }\textbf {\bibinfo {volume} {103}},\ \bibinfo {pages} {4775}
  (\bibinfo {year} {1998})}\BibitemShut {NoStop}%
\bibitem [{\citenamefont {Parashar}\ \emph {et~al.}(2018)\citenamefont
  {Parashar}, \citenamefont {Matthaeus},\ and\ \citenamefont
  {Shay}}]{Parashar2018}%
  \BibitemOpen
  \bibfield  {author} {\bibinfo {author} {\bibfnamefont {T.~N.}\ \bibnamefont
  {Parashar}}, \bibinfo {author} {\bibfnamefont {W.~H.}\ \bibnamefont
  {Matthaeus}}, \ and\ \bibinfo {author} {\bibfnamefont {M.~A.}\ \bibnamefont
  {Shay}},\ }\href {\doibase 10.3847/2041-8213/aadb8b} {\bibfield  {journal}
  {\bibinfo  {journal} {Astrophys. J. L.}\ }\textbf {\bibinfo {volume} {864}},\
  \bibinfo {pages} {L21} (\bibinfo {year} {2018})}\BibitemShut {NoStop}%
\bibitem [{\citenamefont {Leamon}\ \emph {et~al.}(1999)\citenamefont {Leamon},
  \citenamefont {Smith}, \citenamefont {Ness},\ and\ \citenamefont
  {Wong}}]{Leamon1999}%
  \BibitemOpen
  \bibfield  {author} {\bibinfo {author} {\bibfnamefont {R.~J.}\ \bibnamefont
  {Leamon}}, \bibinfo {author} {\bibfnamefont {C.~W.}\ \bibnamefont {Smith}},
  \bibinfo {author} {\bibfnamefont {N.~F.}\ \bibnamefont {Ness}}, \ and\
  \bibinfo {author} {\bibfnamefont {H.~K.}\ \bibnamefont {Wong}},\ }\href
  {\doibase 10.1029/1999JA900158} {\bibfield  {journal} {\bibinfo  {journal}
  {J. Geophys. Res. Sp. Phys.}\ }\textbf {\bibinfo {volume} {104}},\ \bibinfo
  {pages} {22331} (\bibinfo {year} {1999})}\BibitemShut {NoStop}%
\bibitem [{\citenamefont {Boldyrev}\ \emph {et~al.}(2015)\citenamefont
  {Boldyrev}, \citenamefont {Chen}, \citenamefont {Xia},\ and\ \citenamefont
  {Zhdankin}}]{Boldyrev2015}%
  \BibitemOpen
  \bibfield  {author} {\bibinfo {author} {\bibfnamefont {S.}~\bibnamefont
  {Boldyrev}}, \bibinfo {author} {\bibfnamefont {C.~H.~K.}\ \bibnamefont
  {Chen}}, \bibinfo {author} {\bibfnamefont {Q.}~\bibnamefont {Xia}}, \ and\
  \bibinfo {author} {\bibfnamefont {V.}~\bibnamefont {Zhdankin}},\ }\href
  {\doibase 10.1088/0004-637X/806/2/238} {\bibfield  {journal} {\bibinfo
  {journal} {Astrophys. J.}\ }\textbf {\bibinfo {volume} {806}},\ \bibinfo
  {pages} {238} (\bibinfo {year} {2015})}\BibitemShut {NoStop}%
\bibitem [{\citenamefont {Cerri}\ \emph {et~al.}(2019)\citenamefont {Cerri},
  \citenamefont {Gro{\v{s}}elj},\ and\ \citenamefont {Franci}}]{Cerri2019}%
  \BibitemOpen
  \bibfield  {author} {\bibinfo {author} {\bibfnamefont {S.~S.}\ \bibnamefont
  {Cerri}}, \bibinfo {author} {\bibfnamefont {D.}~\bibnamefont
  {Gro{\v{s}}elj}}, \ and\ \bibinfo {author} {\bibfnamefont {L.}~\bibnamefont
  {Franci}},\ }\href {\doibase 10.3389/fspas.2019.00064} {\bibfield  {journal}
  {\bibinfo  {journal} {Frontiers in Astronomy and Space Sciences}\ }\textbf
  {\bibinfo {volume} {6}},\ \bibinfo {pages} {64} (\bibinfo {year}
  {2019})}\BibitemShut {NoStop}%
\bibitem [{\citenamefont {Told}\ \emph {et~al.}(2015)\citenamefont {Told},
  \citenamefont {Jenko}, \citenamefont {TenBarge}, \citenamefont {Howes},\ and\
  \citenamefont {Hammett}}]{Told2015}%
  \BibitemOpen
  \bibfield  {author} {\bibinfo {author} {\bibfnamefont {D.}~\bibnamefont
  {Told}}, \bibinfo {author} {\bibfnamefont {F.}~\bibnamefont {Jenko}},
  \bibinfo {author} {\bibfnamefont {J.~M.}\ \bibnamefont {TenBarge}}, \bibinfo
  {author} {\bibfnamefont {G.~G.}\ \bibnamefont {Howes}}, \ and\ \bibinfo
  {author} {\bibfnamefont {G.~W.}\ \bibnamefont {Hammett}},\ }\href {\doibase
  10.1103/PhysRevLett.115.025003} {\bibfield  {journal} {\bibinfo  {journal}
  {Phys. Rev. Lett.}\ }\textbf {\bibinfo {volume} {115}},\ \bibinfo {pages}
  {025003} (\bibinfo {year} {2015})}\BibitemShut {NoStop}%
\bibitem [{\citenamefont {Gary}(2015)}]{Gary2015}%
  \BibitemOpen
  \bibfield  {author} {\bibinfo {author} {\bibfnamefont {S.~P.}\ \bibnamefont
  {Gary}},\ }\href {\doibase 10.1098/rsta.2014.0149} {\bibfield  {journal}
  {\bibinfo  {journal} {Philos. Trans. R. Soc. A Math. Phys. Eng. Sci.}\
  }\textbf {\bibinfo {volume} {373}},\ \bibinfo {pages} {20140149} (\bibinfo
  {year} {2015})}\BibitemShut {NoStop}%
\bibitem [{\citenamefont {Kuznetsova}\ \emph {et~al.}(1998)\citenamefont
  {Kuznetsova}, \citenamefont {Hesse},\ and\ \citenamefont
  {Winske}}]{Kuznetsova1998}%
  \BibitemOpen
  \bibfield  {author} {\bibinfo {author} {\bibfnamefont {M.~M.}\ \bibnamefont
  {Kuznetsova}}, \bibinfo {author} {\bibfnamefont {M.}~\bibnamefont {Hesse}}, \
  and\ \bibinfo {author} {\bibfnamefont {D.}~\bibnamefont {Winske}},\ }\href
  {\doibase 10.1029/97JA02699} {\bibfield  {journal} {\bibinfo  {journal} {J.
  Geophys. Res.}\ }\textbf {\bibinfo {volume} {103}},\ \bibinfo {pages} {199}
  (\bibinfo {year} {1998})}\BibitemShut {NoStop}%
\bibitem [{\citenamefont {Omura}\ \emph {et~al.}(2008)\citenamefont {Omura},
  \citenamefont {Katoh},\ and\ \citenamefont {Summers}}]{Omura2008}%
  \BibitemOpen
  \bibfield  {author} {\bibinfo {author} {\bibfnamefont {Y.}~\bibnamefont
  {Omura}}, \bibinfo {author} {\bibfnamefont {Y.}~\bibnamefont {Katoh}}, \ and\
  \bibinfo {author} {\bibfnamefont {D.}~\bibnamefont {Summers}},\ }\href
  {\doibase 10.1029/2007JA012622} {\bibfield  {journal} {\bibinfo  {journal}
  {Journal of Geophysical Research: Space Physics}\ }\textbf {\bibinfo {volume}
  {113}},\ \bibinfo {pages} {A04223} (\bibinfo {year} {2008})}\BibitemShut
  {NoStop}%
\bibitem [{\citenamefont {Alexandrova}\ \emph {et~al.}(2009)\citenamefont
  {Alexandrova}, \citenamefont {Saur}, \citenamefont {Lacombe}, \citenamefont
  {Mangeney}, \citenamefont {Mitchell}, \citenamefont {Schwartz},\ and\
  \citenamefont {Robert}}]{Alexandrova2009}%
  \BibitemOpen
  \bibfield  {author} {\bibinfo {author} {\bibfnamefont {O.}~\bibnamefont
  {Alexandrova}}, \bibinfo {author} {\bibfnamefont {J.}~\bibnamefont {Saur}},
  \bibinfo {author} {\bibfnamefont {C.}~\bibnamefont {Lacombe}}, \bibinfo
  {author} {\bibfnamefont {A.}~\bibnamefont {Mangeney}}, \bibinfo {author}
  {\bibfnamefont {J.}~\bibnamefont {Mitchell}}, \bibinfo {author}
  {\bibfnamefont {S.~J.}\ \bibnamefont {Schwartz}}, \ and\ \bibinfo {author}
  {\bibfnamefont {P.}~\bibnamefont {Robert}},\ }\href {\doibase
  10.1103/PhysRevLett.103.165003} {\bibfield  {journal} {\bibinfo  {journal}
  {Phys. Rev. Lett.}\ }\textbf {\bibinfo {volume} {103}},\ \bibinfo {pages}
  {165003} (\bibinfo {year} {2009})}\BibitemShut {NoStop}%
\bibitem [{\citenamefont {Sahraoui}\ \emph {et~al.}(2013)\citenamefont
  {Sahraoui}, \citenamefont {Huang}, \citenamefont {Belmont}, \citenamefont
  {Goldstein}, \citenamefont {R{\'{e}}tino}, \citenamefont {Robert},\ and\
  \citenamefont {{De Patoul}}}]{Sahraoui2013a}%
  \BibitemOpen
  \bibfield  {author} {\bibinfo {author} {\bibfnamefont {F.}~\bibnamefont
  {Sahraoui}}, \bibinfo {author} {\bibfnamefont {S.~Y.}\ \bibnamefont {Huang}},
  \bibinfo {author} {\bibfnamefont {G.}~\bibnamefont {Belmont}}, \bibinfo
  {author} {\bibfnamefont {M.~L.}\ \bibnamefont {Goldstein}}, \bibinfo {author}
  {\bibfnamefont {A.}~\bibnamefont {R{\'{e}}tino}}, \bibinfo {author}
  {\bibfnamefont {P.}~\bibnamefont {Robert}}, \ and\ \bibinfo {author}
  {\bibfnamefont {J.}~\bibnamefont {{De Patoul}}},\ }\href {\doibase
  10.1088/0004-637X/777/1/15} {\bibfield  {journal} {\bibinfo  {journal}
  {Astrophys. J.}\ }\textbf {\bibinfo {volume} {777}},\ \bibinfo {pages} {15}
  (\bibinfo {year} {2013})}\BibitemShut {NoStop}%
\bibitem [{\citenamefont {Huang}\ \emph {et~al.}(2014)\citenamefont {Huang},
  \citenamefont {Sahraoui}, \citenamefont {Deng}, \citenamefont {He},
  \citenamefont {Yuan}, \citenamefont {Zhou}, \citenamefont {Pang},\ and\
  \citenamefont {Fu}}]{Huang2014e}%
  \BibitemOpen
  \bibfield  {author} {\bibinfo {author} {\bibfnamefont {S.~Y.}\ \bibnamefont
  {Huang}}, \bibinfo {author} {\bibfnamefont {F.}~\bibnamefont {Sahraoui}},
  \bibinfo {author} {\bibfnamefont {X.~H.}\ \bibnamefont {Deng}}, \bibinfo
  {author} {\bibfnamefont {J.~S.}\ \bibnamefont {He}}, \bibinfo {author}
  {\bibfnamefont {Z.~G.}\ \bibnamefont {Yuan}}, \bibinfo {author}
  {\bibfnamefont {M.}~\bibnamefont {Zhou}}, \bibinfo {author} {\bibfnamefont
  {Y.}~\bibnamefont {Pang}}, \ and\ \bibinfo {author} {\bibfnamefont {H.~S.}\
  \bibnamefont {Fu}},\ }\href {\doibase 10.1088/2041-8205/789/2/L28} {\bibfield
   {journal} {\bibinfo  {journal} {Astrophys. J.}\ }\textbf {\bibinfo {volume}
  {789}},\ \bibinfo {pages} {L28} (\bibinfo {year} {2014})}\BibitemShut
  {NoStop}%
\bibitem [{\citenamefont {Comisso}\ and\ \citenamefont
  {Sironi}(2022)}]{Comisso2022}%
  \BibitemOpen
  \bibfield  {author} {\bibinfo {author} {\bibfnamefont {L.}~\bibnamefont
  {Comisso}}\ and\ \bibinfo {author} {\bibfnamefont {L.}~\bibnamefont
  {Sironi}},\ }\href {\doibase 10.3847/2041-8213/ac8422} {\bibfield  {journal}
  {\bibinfo  {journal} {The Astrophysical Journal Letters}\ }\textbf {\bibinfo
  {volume} {936}},\ \bibinfo {pages} {L27} (\bibinfo {year}
  {2022})}\BibitemShut {NoStop}%
\bibitem [{\citenamefont {Ricci}\ \emph {et~al.}(2004)\citenamefont {Ricci},
  \citenamefont {Brackbill}, \citenamefont {Daughton},\ and\ \citenamefont
  {Lapenta}}]{Ricci2004}%
  \BibitemOpen
  \bibfield  {author} {\bibinfo {author} {\bibfnamefont {P.}~\bibnamefont
  {Ricci}}, \bibinfo {author} {\bibfnamefont {J.~U.}\ \bibnamefont
  {Brackbill}}, \bibinfo {author} {\bibfnamefont {W.}~\bibnamefont {Daughton}},
  \ and\ \bibinfo {author} {\bibfnamefont {G.}~\bibnamefont {Lapenta}},\ }\href
  {\doibase 10.1063/1.1768552} {\bibfield  {journal} {\bibinfo  {journal}
  {Phys. Plasmas}\ }\textbf {\bibinfo {volume} {11}},\ \bibinfo {pages} {4102}
  (\bibinfo {year} {2004})}\BibitemShut {NoStop}%
\bibitem [{\citenamefont {Hesse}\ \emph {et~al.}(2004)\citenamefont {Hesse},
  \citenamefont {Kuznetsova},\ and\ \citenamefont {Birn}}]{Hesse2004}%
  \BibitemOpen
  \bibfield  {author} {\bibinfo {author} {\bibfnamefont {M.}~\bibnamefont
  {Hesse}}, \bibinfo {author} {\bibfnamefont {M.}~\bibnamefont {Kuznetsova}}, \
  and\ \bibinfo {author} {\bibfnamefont {J.}~\bibnamefont {Birn}},\ }\href
  {\doibase 10.1063/1.1795991} {\bibfield  {journal} {\bibinfo  {journal}
  {Phys. Plasmas}\ }\textbf {\bibinfo {volume} {11}},\ \bibinfo {pages} {5387}
  (\bibinfo {year} {2004})}\BibitemShut {NoStop}%
\bibitem [{\citenamefont {Azizabadi}\ \emph {et~al.}(2021)\citenamefont
  {Azizabadi}, \citenamefont {Jain},\ and\ \citenamefont
  {Büchner}}]{Azizabadi2021}%
  \BibitemOpen
  \bibfield  {author} {\bibinfo {author} {\bibfnamefont {A.~C.}\ \bibnamefont
  {Azizabadi}}, \bibinfo {author} {\bibfnamefont {N.}~\bibnamefont {Jain}}, \
  and\ \bibinfo {author} {\bibfnamefont {J.}~\bibnamefont {Büchner}},\ }\href
  {\doibase 10.1063/5.0040692} {\bibfield  {journal} {\bibinfo  {journal}
  {Physics of Plasmas}\ }\textbf {\bibinfo {volume} {28}},\ \bibinfo {pages}
  {052904} (\bibinfo {year} {2021})}\BibitemShut {NoStop}%
\bibitem [{\citenamefont {Jain}\ \emph {et~al.}(2022)\citenamefont {Jain},
  \citenamefont {Mu{\~{n}}oz}, \citenamefont {Tabriz}, \citenamefont {Rampp},\
  and\ \citenamefont {Büchner}}]{Jain2022}%
  \BibitemOpen
  \bibfield  {author} {\bibinfo {author} {\bibfnamefont {N.}~\bibnamefont
  {Jain}}, \bibinfo {author} {\bibfnamefont {P.~A.}\ \bibnamefont
  {Mu{\~{n}}oz}}, \bibinfo {author} {\bibfnamefont {M.~F.}\ \bibnamefont
  {Tabriz}}, \bibinfo {author} {\bibfnamefont {M.}~\bibnamefont {Rampp}}, \
  and\ \bibinfo {author} {\bibfnamefont {J.}~\bibnamefont {Büchner}},\ }\href
  {\doibase 10.1063/5.0087103} {\bibfield  {journal} {\bibinfo  {journal}
  {Physics of Plasmas}\ }\textbf {\bibinfo {volume} {29}},\ \bibinfo {eid}
  {053902} (\bibinfo {year} {2022})}\BibitemShut {NoStop}%
\bibitem [{\citenamefont {Jain}\ and\ \citenamefont
  {B{\"{u}}chner}(2014)}]{Jain2014c}%
  \BibitemOpen
  \bibfield  {author} {\bibinfo {author} {\bibfnamefont {N.}~\bibnamefont
  {Jain}}\ and\ \bibinfo {author} {\bibfnamefont {J.}~\bibnamefont
  {B{\"{u}}chner}},\ }\href {\doibase 10.1063/1.4887279} {\bibfield  {journal}
  {\bibinfo  {journal} {Phys. Plasmas}\ }\textbf {\bibinfo {volume} {21}},\
  \bibinfo {pages} {072306} (\bibinfo {year} {2014})}\BibitemShut {NoStop}%
\bibitem [{\citenamefont {Jain}\ and\ \citenamefont
  {B{\"{u}}chner}(2015)}]{Jain2015c}%
  \BibitemOpen
  \bibfield  {author} {\bibinfo {author} {\bibfnamefont {N.}~\bibnamefont
  {Jain}}\ and\ \bibinfo {author} {\bibfnamefont {J.}~\bibnamefont
  {B{\"{u}}chner}},\ }\href {\doibase 10.1017/S0022377815001257} {\bibfield
  {journal} {\bibinfo  {journal} {J. Plasma Phys.}\ }\textbf {\bibinfo {volume}
  {81}},\ \bibinfo {pages} {905810606} (\bibinfo {year} {2015})}\BibitemShut
  {NoStop}%
\bibitem [{\citenamefont {Phan}\ \emph {et~al.}(2018)\citenamefont {Phan},
  \citenamefont {Eastwood}, \citenamefont {Shay}, \citenamefont {Drake},
  \citenamefont {Sonnerup}, \citenamefont {Fujimoto}, \citenamefont {Cassak},
  \citenamefont {{\O}ieroset}, \citenamefont {Burch}, \citenamefont {Torbert},
  \citenamefont {Rager}, \citenamefont {Dorelli}, \citenamefont {Gershman},
  \citenamefont {Pollock}, \citenamefont {Pyakurel}, \citenamefont {Haggerty},
  \citenamefont {Khotyaintsev}, \citenamefont {Lavraud}, \citenamefont {Saito},
  \citenamefont {Oka}, \citenamefont {Ergun}, \citenamefont {Retino},
  \citenamefont {{Le Contel}}, \citenamefont {Argall}, \citenamefont {Giles},
  \citenamefont {Moore}, \citenamefont {Wilder}, \citenamefont {Strangeway},
  \citenamefont {Russell}, \citenamefont {Lindqvist},\ and\ \citenamefont
  {Magnes}}]{Phan2018}%
  \BibitemOpen
  \bibfield  {author} {\bibinfo {author} {\bibfnamefont {T.~D.}\ \bibnamefont
  {Phan}}, \bibinfo {author} {\bibfnamefont {J.~P.}\ \bibnamefont {Eastwood}},
  \bibinfo {author} {\bibfnamefont {M.~A.}\ \bibnamefont {Shay}}, \bibinfo
  {author} {\bibfnamefont {J.~F.}\ \bibnamefont {Drake}}, \bibinfo {author}
  {\bibfnamefont {B.~U.~{\"{O}}.}\ \bibnamefont {Sonnerup}}, \bibinfo {author}
  {\bibfnamefont {M.}~\bibnamefont {Fujimoto}}, \bibinfo {author}
  {\bibfnamefont {P.~A.}\ \bibnamefont {Cassak}}, \bibinfo {author}
  {\bibfnamefont {M.}~\bibnamefont {{\O}ieroset}}, \bibinfo {author}
  {\bibfnamefont {J.~L.}\ \bibnamefont {Burch}}, \bibinfo {author}
  {\bibfnamefont {R.~B.}\ \bibnamefont {Torbert}}, \bibinfo {author}
  {\bibfnamefont {A.~C.}\ \bibnamefont {Rager}}, \bibinfo {author}
  {\bibfnamefont {J.~C.}\ \bibnamefont {Dorelli}}, \bibinfo {author}
  {\bibfnamefont {D.~J.}\ \bibnamefont {Gershman}}, \bibinfo {author}
  {\bibfnamefont {C.}~\bibnamefont {Pollock}}, \bibinfo {author} {\bibfnamefont
  {P.~S.}\ \bibnamefont {Pyakurel}}, \bibinfo {author} {\bibfnamefont {C.~C.}\
  \bibnamefont {Haggerty}}, \bibinfo {author} {\bibfnamefont {Y.}~\bibnamefont
  {Khotyaintsev}}, \bibinfo {author} {\bibfnamefont {B.}~\bibnamefont
  {Lavraud}}, \bibinfo {author} {\bibfnamefont {Y.}~\bibnamefont {Saito}},
  \bibinfo {author} {\bibfnamefont {M.}~\bibnamefont {Oka}}, \bibinfo {author}
  {\bibfnamefont {R.~E.}\ \bibnamefont {Ergun}}, \bibinfo {author}
  {\bibfnamefont {A.}~\bibnamefont {Retino}}, \bibinfo {author} {\bibfnamefont
  {O.}~\bibnamefont {{Le Contel}}}, \bibinfo {author} {\bibfnamefont {M.~R.}\
  \bibnamefont {Argall}}, \bibinfo {author} {\bibfnamefont {B.~L.}\
  \bibnamefont {Giles}}, \bibinfo {author} {\bibfnamefont {T.~E.}\ \bibnamefont
  {Moore}}, \bibinfo {author} {\bibfnamefont {F.~D.}\ \bibnamefont {Wilder}},
  \bibinfo {author} {\bibfnamefont {R.~J.}\ \bibnamefont {Strangeway}},
  \bibinfo {author} {\bibfnamefont {C.~T.}\ \bibnamefont {Russell}}, \bibinfo
  {author} {\bibfnamefont {P.~A.}\ \bibnamefont {Lindqvist}}, \ and\ \bibinfo
  {author} {\bibfnamefont {W.}~\bibnamefont {Magnes}},\ }\href {\doibase
  10.1038/s41586-018-0091-5} {\bibfield  {journal} {\bibinfo  {journal}
  {Nature}\ }\textbf {\bibinfo {volume} {557}},\ \bibinfo {pages} {202}
  (\bibinfo {year} {2018})}\BibitemShut {NoStop}%
\bibitem [{\citenamefont {Mu{\~{n}}oz}\ \emph {et~al.}(2018)\citenamefont
  {Mu{\~{n}}oz}, \citenamefont {Jain}, \citenamefont {Kilian},\ and\
  \citenamefont {B{\"{u}}chner}}]{Munoz2018}%
  \BibitemOpen
  \bibfield  {author} {\bibinfo {author} {\bibfnamefont {P.~A.}\ \bibnamefont
  {Mu{\~{n}}oz}}, \bibinfo {author} {\bibfnamefont {N.}~\bibnamefont {Jain}},
  \bibinfo {author} {\bibfnamefont {P.}~\bibnamefont {Kilian}}, \ and\ \bibinfo
  {author} {\bibfnamefont {J.}~\bibnamefont {B{\"{u}}chner}},\ }\href {\doibase
  10.1016/j.cpc.2017.10.012} {\bibfield  {journal} {\bibinfo  {journal}
  {Comput. Phys. Commun.}\ }\textbf {\bibinfo {volume} {224}},\ \bibinfo
  {pages} {245} (\bibinfo {year} {2018})}\BibitemShut {NoStop}%
\bibitem [{\citenamefont {Forslund}\ and\ \citenamefont
  {Freidberg}(1971)}]{Forslund1971}%
  \BibitemOpen
  \bibfield  {author} {\bibinfo {author} {\bibfnamefont {D.~W.}\ \bibnamefont
  {Forslund}}\ and\ \bibinfo {author} {\bibfnamefont {J.~P.}\ \bibnamefont
  {Freidberg}},\ }\href {\doibase 10.1103/PhysRevLett.27.1189} {\bibfield
  {journal} {\bibinfo  {journal} {Phys. Rev. Lett.}\ }\textbf {\bibinfo
  {volume} {27}},\ \bibinfo {pages} {1189} (\bibinfo {year}
  {1971})}\BibitemShut {NoStop}%
\bibitem [{\citenamefont {Hewett}\ and\ \citenamefont
  {Nielson}(1978)}]{Hewett1978}%
  \BibitemOpen
  \bibfield  {author} {\bibinfo {author} {\bibfnamefont {D.}~\bibnamefont
  {Hewett}}\ and\ \bibinfo {author} {\bibfnamefont {C.}~\bibnamefont
  {Nielson}},\ }\href {\doibase 10.1016/0021-9991(78)90153-5} {\bibfield
  {journal} {\bibinfo  {journal} {J. Comput. Phys.}\ }\textbf {\bibinfo
  {volume} {29}},\ \bibinfo {pages} {219} (\bibinfo {year} {1978})}\BibitemShut
  {NoStop}%
\bibitem [{\citenamefont {Swift}(1996)}]{Swift1996}%
  \BibitemOpen
  \bibfield  {author} {\bibinfo {author} {\bibfnamefont {D.~W.}\ \bibnamefont
  {Swift}},\ }\href {\doibase 10.1006/jcph.1996.0124} {\bibfield  {journal}
  {\bibinfo  {journal} {J. Comput. Phys.}\ }\textbf {\bibinfo {volume} {126}},\
  \bibinfo {pages} {109} (\bibinfo {year} {1996})}\BibitemShut {NoStop}%
\bibitem [{\citenamefont {Shay}\ \emph {et~al.}(1998)\citenamefont {Shay},
  \citenamefont {Drake}, \citenamefont {Denton},\ and\ \citenamefont
  {Biskamp}}]{Shay1998}%
  \BibitemOpen
  \bibfield  {author} {\bibinfo {author} {\bibfnamefont {M.~A.}\ \bibnamefont
  {Shay}}, \bibinfo {author} {\bibfnamefont {J.~F.}\ \bibnamefont {Drake}},
  \bibinfo {author} {\bibfnamefont {R.~E.}\ \bibnamefont {Denton}}, \ and\
  \bibinfo {author} {\bibfnamefont {D.}~\bibnamefont {Biskamp}},\ }\href
  {\doibase 10.1029/97JA03528} {\bibfield  {journal} {\bibinfo  {journal} {J.
  Geophys. Res.}\ }\textbf {\bibinfo {volume} {103}},\ \bibinfo {pages} {9165}
  (\bibinfo {year} {1998})}\BibitemShut {NoStop}%
\bibitem [{\citenamefont {Valentini}\ \emph {et~al.}(2007)\citenamefont
  {Valentini}, \citenamefont {Tr{\'{a}}vn{\'{i}}{\v{c}}ek}, \citenamefont
  {Califano}, \citenamefont {Hellinger},\ and\ \citenamefont
  {Mangeney}}]{Valentini2007}%
  \BibitemOpen
  \bibfield  {author} {\bibinfo {author} {\bibfnamefont {F.}~\bibnamefont
  {Valentini}}, \bibinfo {author} {\bibfnamefont {P.}~\bibnamefont
  {Tr{\'{a}}vn{\'{i}}{\v{c}}ek}}, \bibinfo {author} {\bibfnamefont
  {F.}~\bibnamefont {Califano}}, \bibinfo {author} {\bibfnamefont
  {P.}~\bibnamefont {Hellinger}}, \ and\ \bibinfo {author} {\bibfnamefont
  {A.}~\bibnamefont {Mangeney}},\ }\href {\doibase 10.1016/j.jcp.2007.01.001}
  {\bibfield  {journal} {\bibinfo  {journal} {J. Comput. Phys.}\ }\textbf
  {\bibinfo {volume} {225}},\ \bibinfo {pages} {753} (\bibinfo {year}
  {2007})}\BibitemShut {NoStop}%
\bibitem [{\citenamefont {Amano}\ \emph {et~al.}(2014)\citenamefont {Amano},
  \citenamefont {Higashimori},\ and\ \citenamefont {Shirakawa}}]{Amano2014}%
  \BibitemOpen
  \bibfield  {author} {\bibinfo {author} {\bibfnamefont {T.}~\bibnamefont
  {Amano}}, \bibinfo {author} {\bibfnamefont {K.}~\bibnamefont {Higashimori}},
  \ and\ \bibinfo {author} {\bibfnamefont {K.}~\bibnamefont {Shirakawa}},\
  }\href {\doibase 10.1016/j.jcp.2014.06.048} {\bibfield  {journal} {\bibinfo
  {journal} {J. Comput. Phys.}\ }\textbf {\bibinfo {volume} {275}},\ \bibinfo
  {pages} {197} (\bibinfo {year} {2014})}\BibitemShut {NoStop}%
\bibitem [{\citenamefont {Howes}(2015{\natexlab{b}})}]{Howes2014b}%
  \BibitemOpen
  \bibfield  {author} {\bibinfo {author} {\bibfnamefont {G.~G.}\ \bibnamefont
  {Howes}},\ }\href {\doibase 10.1017/S0022377814001056} {\bibfield  {journal}
  {\bibinfo  {journal} {J. Plasma Phys.}\ }\textbf {\bibinfo {volume} {81}},\
  \bibinfo {pages} {325810203} (\bibinfo {year}
  {2015}{\natexlab{b}})}\BibitemShut {NoStop}%
\bibitem [{\citenamefont {Li}\ \emph {et~al.}(2016)\citenamefont {Li},
  \citenamefont {Howes}, \citenamefont {Klein},\ and\ \citenamefont
  {TenBarge}}]{Li2016bn}%
  \BibitemOpen
  \bibfield  {author} {\bibinfo {author} {\bibfnamefont {T.~C.}\ \bibnamefont
  {Li}}, \bibinfo {author} {\bibfnamefont {G.~G.}\ \bibnamefont {Howes}},
  \bibinfo {author} {\bibfnamefont {K.~G.}\ \bibnamefont {Klein}}, \ and\
  \bibinfo {author} {\bibfnamefont {J.~M.}\ \bibnamefont {TenBarge}},\ }\href
  {\doibase 10.3847/2041-8205/832/2/L24} {\bibfield  {journal} {\bibinfo
  {journal} {Astrophys. J.}\ }\textbf {\bibinfo {volume} {832}},\ \bibinfo
  {pages} {L24} (\bibinfo {year} {2016})}\BibitemShut {NoStop}%
\bibitem [{\citenamefont {Franci}\ \emph {et~al.}(2018)\citenamefont {Franci},
  \citenamefont {Landi}, \citenamefont {Verdini}, \citenamefont {Matteini},\
  and\ \citenamefont {Hellinger}}]{Franci2018b}%
  \BibitemOpen
  \bibfield  {author} {\bibinfo {author} {\bibfnamefont {L.}~\bibnamefont
  {Franci}}, \bibinfo {author} {\bibfnamefont {S.}~\bibnamefont {Landi}},
  \bibinfo {author} {\bibfnamefont {A.}~\bibnamefont {Verdini}}, \bibinfo
  {author} {\bibfnamefont {L.}~\bibnamefont {Matteini}}, \ and\ \bibinfo
  {author} {\bibfnamefont {P.}~\bibnamefont {Hellinger}},\ }\href {\doibase
  10.3847/1538-4357/aaa3e8} {\bibfield  {journal} {\bibinfo  {journal}
  {Astrophys. J.}\ }\textbf {\bibinfo {volume} {853}},\ \bibinfo {pages} {26}
  (\bibinfo {year} {2018})}\BibitemShut {NoStop}%
\bibitem [{\citenamefont {Gary}\ \emph {et~al.}(2020)\citenamefont {Gary},
  \citenamefont {Bandyopadhyay}, \citenamefont {Qudsi}, \citenamefont
  {Matthaeus}, \citenamefont {Maruca}, \citenamefont {Parashar},\ and\
  \citenamefont {Roytershteyn}}]{Gary2020}%
  \BibitemOpen
  \bibfield  {author} {\bibinfo {author} {\bibfnamefont {S.~P.}\ \bibnamefont
  {Gary}}, \bibinfo {author} {\bibfnamefont {R.}~\bibnamefont {Bandyopadhyay}},
  \bibinfo {author} {\bibfnamefont {R.~A.}\ \bibnamefont {Qudsi}}, \bibinfo
  {author} {\bibfnamefont {W.~H.}\ \bibnamefont {Matthaeus}}, \bibinfo {author}
  {\bibfnamefont {B.~A.}\ \bibnamefont {Maruca}}, \bibinfo {author}
  {\bibfnamefont {T.~N.}\ \bibnamefont {Parashar}}, \ and\ \bibinfo {author}
  {\bibfnamefont {V.}~\bibnamefont {Roytershteyn}},\ }\href {\doibase
  10.3847/1538-4357/abb2ac} {\bibfield  {journal} {\bibinfo  {journal} {The
  Astrophysical Journal}\ }\textbf {\bibinfo {volume} {901}},\ \bibinfo {pages}
  {160} (\bibinfo {year} {2020})}\BibitemShut {NoStop}%
\bibitem [{\citenamefont {B{\"{u}}chner}\ and\ \citenamefont
  {Kuska}(1997)}]{Buchner1997}%
  \BibitemOpen
  \bibfield  {author} {\bibinfo {author} {\bibfnamefont {J.}~\bibnamefont
  {B{\"{u}}chner}}\ and\ \bibinfo {author} {\bibfnamefont {J.~P.}\ \bibnamefont
  {Kuska}},\ }\href {\doibase 10.1016/S0273-1177(97)00082-3} {\bibfield
  {journal} {\bibinfo  {journal} {Adv. Sp. Res.}\ }\textbf {\bibinfo {volume}
  {19}},\ \bibinfo {pages} {1817} (\bibinfo {year} {1997})}\BibitemShut
  {NoStop}%
\bibitem [{\citenamefont {Horiuchi}\ and\ \citenamefont
  {Sato}(1999)}]{Horiuchi1999a}%
  \BibitemOpen
  \bibfield  {author} {\bibinfo {author} {\bibfnamefont {R.}~\bibnamefont
  {Horiuchi}}\ and\ \bibinfo {author} {\bibfnamefont {T.}~\bibnamefont
  {Sato}},\ }\href {\doibase 10.1063/1.873744} {\bibfield  {journal} {\bibinfo
  {journal} {Physics of Plasmas}\ }\textbf {\bibinfo {volume} {6}},\ \bibinfo
  {pages} {4565} (\bibinfo {year} {1999})}\BibitemShut {NoStop}%
\bibitem [{\citenamefont {Fujimoto}(2011)}]{Fujimoto2011a}%
  \BibitemOpen
  \bibfield  {author} {\bibinfo {author} {\bibfnamefont {K.}~\bibnamefont
  {Fujimoto}},\ }\href {\doibase 10.1063/1.3642609} {\bibfield  {journal}
  {\bibinfo  {journal} {Phys. Plasmas}\ }\textbf {\bibinfo {volume} {18}},\
  \bibinfo {pages} {111206} (\bibinfo {year} {2011})}\BibitemShut {NoStop}%
\bibitem [{\citenamefont {Che}\ \emph {et~al.}(2011)\citenamefont {Che},
  \citenamefont {Drake},\ and\ \citenamefont {Swisdak}}]{Che2011}%
  \BibitemOpen
  \bibfield  {author} {\bibinfo {author} {\bibfnamefont {H.}~\bibnamefont
  {Che}}, \bibinfo {author} {\bibfnamefont {J.~F.}\ \bibnamefont {Drake}}, \
  and\ \bibinfo {author} {\bibfnamefont {M.}~\bibnamefont {Swisdak}},\ }\href
  {\doibase 10.1038/nature10091} {\bibfield  {journal} {\bibinfo  {journal}
  {Nature}\ }\textbf {\bibinfo {volume} {474}},\ \bibinfo {pages} {184}
  (\bibinfo {year} {2011})}\BibitemShut {NoStop}%
\bibitem [{\citenamefont {Daughton}\ \emph {et~al.}(2011)\citenamefont
  {Daughton}, \citenamefont {Roytershteyn}, \citenamefont {Karimabadi},
  \citenamefont {Yin}, \citenamefont {Albright}, \citenamefont {Bergen},\ and\
  \citenamefont {Bowers}}]{Daughton2011}%
  \BibitemOpen
  \bibfield  {author} {\bibinfo {author} {\bibfnamefont {W.}~\bibnamefont
  {Daughton}}, \bibinfo {author} {\bibfnamefont {V.}~\bibnamefont
  {Roytershteyn}}, \bibinfo {author} {\bibfnamefont {H.}~\bibnamefont
  {Karimabadi}}, \bibinfo {author} {\bibfnamefont {L.}~\bibnamefont {Yin}},
  \bibinfo {author} {\bibfnamefont {B.~J.}\ \bibnamefont {Albright}}, \bibinfo
  {author} {\bibfnamefont {B.}~\bibnamefont {Bergen}}, \ and\ \bibinfo {author}
  {\bibfnamefont {K.~J.}\ \bibnamefont {Bowers}},\ }\href {\doibase
  10.1038/nphys1965} {\bibfield  {journal} {\bibinfo  {journal} {Nat. Phys.}\
  }\textbf {\bibinfo {volume} {7}},\ \bibinfo {pages} {539} (\bibinfo {year}
  {2011})}\BibitemShut {NoStop}%
\bibitem [{\citenamefont {Cerri}\ \emph {et~al.}(2017)\citenamefont {Cerri},
  \citenamefont {Servidio},\ and\ \citenamefont {Califano}}]{Cerri2017a}%
  \BibitemOpen
  \bibfield  {author} {\bibinfo {author} {\bibfnamefont {S.~S.}\ \bibnamefont
  {Cerri}}, \bibinfo {author} {\bibfnamefont {S.}~\bibnamefont {Servidio}}, \
  and\ \bibinfo {author} {\bibfnamefont {F.}~\bibnamefont {Califano}},\ }\href
  {\doibase 10.3847/2041-8213/aa87b0} {\bibfield  {journal} {\bibinfo
  {journal} {Astrophys. J.}\ }\textbf {\bibinfo {volume} {846}},\ \bibinfo
  {pages} {L18} (\bibinfo {year} {2017})}\BibitemShut {NoStop}%
\bibitem [{\citenamefont {Sisti}\ \emph {et~al.}(2021)\citenamefont {Sisti},
  \citenamefont {Fadanelli}, \citenamefont {Cerri}, \citenamefont {Faganello},
  \citenamefont {Califano},\ and\ \citenamefont {Agullo}}]{Sisti2021}%
  \BibitemOpen
  \bibfield  {author} {\bibinfo {author} {\bibfnamefont {M.}~\bibnamefont
  {Sisti}}, \bibinfo {author} {\bibfnamefont {S.}~\bibnamefont {Fadanelli}},
  \bibinfo {author} {\bibfnamefont {S.~S.}\ \bibnamefont {Cerri}}, \bibinfo
  {author} {\bibfnamefont {M.}~\bibnamefont {Faganello}}, \bibinfo {author}
  {\bibfnamefont {F.}~\bibnamefont {Califano}}, \ and\ \bibinfo {author}
  {\bibfnamefont {O.}~\bibnamefont {Agullo}},\ }\href {\doibase
  10.1051/0004-6361/202141902} {\bibfield  {journal} {\bibinfo  {journal}
  {Astronomy and Astrophysics}\ }\textbf {\bibinfo {volume} {655}},\ \bibinfo
  {pages} {A107} (\bibinfo {year} {2021})}\BibitemShut {NoStop}%
\bibitem [{\citenamefont {Rueda}\ \emph {et~al.}(2021)\citenamefont {Rueda},
  \citenamefont {Verscharen}, \citenamefont {Wicks}, \citenamefont {Owen},
  \citenamefont {Nicolaou}, \citenamefont {Walsh}, \citenamefont {Zouganelis},
  \citenamefont {Germaschewski},\ and\ \citenamefont
  {Dom{\'{\i}}nguez}}]{Rueda2021}%
  \BibitemOpen
  \bibfield  {author} {\bibinfo {author} {\bibfnamefont {J.~A.~A.}\
  \bibnamefont {Rueda}}, \bibinfo {author} {\bibfnamefont {D.}~\bibnamefont
  {Verscharen}}, \bibinfo {author} {\bibfnamefont {R.~T.}\ \bibnamefont
  {Wicks}}, \bibinfo {author} {\bibfnamefont {C.~J.}\ \bibnamefont {Owen}},
  \bibinfo {author} {\bibfnamefont {G.}~\bibnamefont {Nicolaou}}, \bibinfo
  {author} {\bibfnamefont {A.~P.}\ \bibnamefont {Walsh}}, \bibinfo {author}
  {\bibfnamefont {I.}~\bibnamefont {Zouganelis}}, \bibinfo {author}
  {\bibfnamefont {K.}~\bibnamefont {Germaschewski}}, \ and\ \bibinfo {author}
  {\bibfnamefont {S.~V.}\ \bibnamefont {Dom{\'{\i}}nguez}},\ }\href {\doibase
  10.1017/S0022377821000404} {\bibfield  {journal} {\bibinfo  {journal}
  {Journal of Plasma Physics}\ }\textbf {\bibinfo {volume} {87}},\ \bibinfo
  {pages} {905870228} (\bibinfo {year} {2021})}\BibitemShut {NoStop}%
\bibitem [{\citenamefont {Franci}\ \emph {et~al.}(2022)\citenamefont {Franci},
  \citenamefont {Papini}, \citenamefont {Micera}, \citenamefont {Lapenta},
  \citenamefont {Hellinger}, \citenamefont {Sarto}, \citenamefont {Burgess},\
  and\ \citenamefont {Landi}}]{Franci2022}%
  \BibitemOpen
  \bibfield  {author} {\bibinfo {author} {\bibfnamefont {L.}~\bibnamefont
  {Franci}}, \bibinfo {author} {\bibfnamefont {E.}~\bibnamefont {Papini}},
  \bibinfo {author} {\bibfnamefont {A.}~\bibnamefont {Micera}}, \bibinfo
  {author} {\bibfnamefont {G.}~\bibnamefont {Lapenta}}, \bibinfo {author}
  {\bibfnamefont {P.}~\bibnamefont {Hellinger}}, \bibinfo {author}
  {\bibfnamefont {D.~D.}\ \bibnamefont {Sarto}}, \bibinfo {author}
  {\bibfnamefont {D.}~\bibnamefont {Burgess}}, \ and\ \bibinfo {author}
  {\bibfnamefont {S.}~\bibnamefont {Landi}},\ }\href {\doibase
  10.3847/1538-4357/ac7da6} {\bibfield  {journal} {\bibinfo  {journal} {The
  Astrophysical Journal}\ }\textbf {\bibinfo {volume} {936}},\ \bibinfo {pages}
  {27} (\bibinfo {year} {2022})}\BibitemShut {NoStop}%
\bibitem [{\citenamefont {Gro{\v{s}}elj}\ \emph {et~al.}(2017)\citenamefont
  {Gro{\v{s}}elj}, \citenamefont {Cerri}, \citenamefont {Navarro},
  \citenamefont {Willmott}, \citenamefont {Told}, \citenamefont {Loureiro},
  \citenamefont {Califano},\ and\ \citenamefont {Jenko}}]{Groselj2017}%
  \BibitemOpen
  \bibfield  {author} {\bibinfo {author} {\bibfnamefont {D.}~\bibnamefont
  {Gro{\v{s}}elj}}, \bibinfo {author} {\bibfnamefont {S.~S.}\ \bibnamefont
  {Cerri}}, \bibinfo {author} {\bibfnamefont {A.~B.}\ \bibnamefont {Navarro}},
  \bibinfo {author} {\bibfnamefont {C.}~\bibnamefont {Willmott}}, \bibinfo
  {author} {\bibfnamefont {D.}~\bibnamefont {Told}}, \bibinfo {author}
  {\bibfnamefont {N.~F.}\ \bibnamefont {Loureiro}}, \bibinfo {author}
  {\bibfnamefont {F.}~\bibnamefont {Califano}}, \ and\ \bibinfo {author}
  {\bibfnamefont {F.}~\bibnamefont {Jenko}},\ }\href {\doibase
  10.3847/1538-4357/aa894d} {\bibfield  {journal} {\bibinfo  {journal}
  {Astrophys. J.}\ }\textbf {\bibinfo {volume} {847}},\ \bibinfo {pages} {28}
  (\bibinfo {year} {2017})}\BibitemShut {NoStop}%
\bibitem [{\citenamefont {Gonz{\'{a}}lez}\ \emph {et~al.}(2019)\citenamefont
  {Gonz{\'{a}}lez}, \citenamefont {Parashar}, \citenamefont {Gomez},
  \citenamefont {Matthaeus},\ and\ \citenamefont {Dmitruk}}]{Gonzalez2019a}%
  \BibitemOpen
  \bibfield  {author} {\bibinfo {author} {\bibfnamefont {C.~A.}\ \bibnamefont
  {Gonz{\'{a}}lez}}, \bibinfo {author} {\bibfnamefont {T.~N.}\ \bibnamefont
  {Parashar}}, \bibinfo {author} {\bibfnamefont {D.}~\bibnamefont {Gomez}},
  \bibinfo {author} {\bibfnamefont {W.~H.}\ \bibnamefont {Matthaeus}}, \ and\
  \bibinfo {author} {\bibfnamefont {P.}~\bibnamefont {Dmitruk}},\ }\href
  {\doibase 10.1063/1.5054110} {\bibfield  {journal} {\bibinfo  {journal}
  {Phys. Plasmas}\ }\textbf {\bibinfo {volume} {26}},\ \bibinfo {pages}
  {012306} (\bibinfo {year} {2019})}\BibitemShut {NoStop}%
\bibitem [{\citenamefont {Jain}\ \emph {et~al.}(2023)\citenamefont {Jain},
  \citenamefont {Mu{\~{n}}oz},\ and\ \citenamefont {B{\"u}chner}}]{Jain2023}%
  \BibitemOpen
  \bibfield  {author} {\bibinfo {author} {\bibfnamefont {N.}~\bibnamefont
  {Jain}}, \bibinfo {author} {\bibfnamefont {P.~A.}\ \bibnamefont
  {Mu{\~{n}}oz}}, \ and\ \bibinfo {author} {\bibfnamefont {J.}~\bibnamefont
  {B{\"u}chner}},\ }\enquote {\bibinfo {title} {Hybrid--kinetic approach:
  Inertial electrons},}\ in\ \href {\doibase 10.1007/978-3-031-11870-8_9}
  {\emph {\bibinfo {booktitle} {Space and Astrophysical Plasma Simulation}}},\
  \bibinfo {editor} {edited by\ \bibinfo {editor} {\bibfnamefont
  {J.}~\bibnamefont {B{\"u}chner}}}\ (\bibinfo  {publisher} {Springer
  International Publishing},\ \bibinfo {year} {2023})\ pp.\ \bibinfo {pages}
  {283--311}\BibitemShut {NoStop}%
\bibitem [{\citenamefont {Bale}\ \emph {et~al.}(2009)\citenamefont {Bale},
  \citenamefont {Kasper}, \citenamefont {Howes}, \citenamefont {Quataert},
  \citenamefont {Salem},\ and\ \citenamefont {Sundkvist}}]{Bale2009}%
  \BibitemOpen
  \bibfield  {author} {\bibinfo {author} {\bibfnamefont {S.}~\bibnamefont
  {Bale}}, \bibinfo {author} {\bibfnamefont {J.}~\bibnamefont {Kasper}},
  \bibinfo {author} {\bibfnamefont {G.~G.}\ \bibnamefont {Howes}}, \bibinfo
  {author} {\bibfnamefont {E.}~\bibnamefont {Quataert}}, \bibinfo {author}
  {\bibfnamefont {C.}~\bibnamefont {Salem}}, \ and\ \bibinfo {author}
  {\bibfnamefont {D.}~\bibnamefont {Sundkvist}},\ }\href {\doibase
  10.1103/PhysRevLett.103.211101} {\bibfield  {journal} {\bibinfo  {journal}
  {Phys. Rev. Lett.}\ }\textbf {\bibinfo {volume} {103}},\ \bibinfo {pages}
  {211101} (\bibinfo {year} {2009})}\BibitemShut {NoStop}%
\bibitem [{\citenamefont {Le}\ \emph {et~al.}(2016)\citenamefont {Le},
  \citenamefont {Daughton}, \citenamefont {Karimabadi},\ and\ \citenamefont
  {Egedal}}]{Le2016}%
  \BibitemOpen
  \bibfield  {author} {\bibinfo {author} {\bibfnamefont {A.}~\bibnamefont
  {Le}}, \bibinfo {author} {\bibfnamefont {W.}~\bibnamefont {Daughton}},
  \bibinfo {author} {\bibfnamefont {H.}~\bibnamefont {Karimabadi}}, \ and\
  \bibinfo {author} {\bibfnamefont {J.}~\bibnamefont {Egedal}},\ }\href
  {\doibase 10.1063/1.4943893} {\bibfield  {journal} {\bibinfo  {journal}
  {Phys. Plasmas}\ }\textbf {\bibinfo {volume} {23}},\ \bibinfo {pages}
  {032114} (\bibinfo {year} {2016})}\BibitemShut {NoStop}%
\bibitem [{\citenamefont {Hesse}\ \emph {et~al.}(1995)\citenamefont {Hesse},
  \citenamefont {Winske},\ and\ \citenamefont {Kuznetsova}}]{Hesse1995}%
  \BibitemOpen
  \bibfield  {author} {\bibinfo {author} {\bibfnamefont {M.}~\bibnamefont
  {Hesse}}, \bibinfo {author} {\bibfnamefont {D.}~\bibnamefont {Winske}}, \
  and\ \bibinfo {author} {\bibfnamefont {M.~M.}\ \bibnamefont {Kuznetsova}},\
  }\href {\doibase 10.1029/95JA01559} {\bibfield  {journal} {\bibinfo
  {journal} {J. Geophys. Res. Sp. Phys.}\ }\textbf {\bibinfo {volume} {100}},\
  \bibinfo {pages} {21815} (\bibinfo {year} {1995})}\BibitemShut {NoStop}%
\bibitem [{\citenamefont {{Max Planck Computing and Data
  Facility}}(2023)}]{MPCDF_Raven_doc}%
  \BibitemOpen
  \bibfield  {author} {\bibinfo {author} {\bibnamefont {{Max Planck Computing
  and Data Facility}}},\ }\href
  {https://docs.mpcdf.mpg.de/doc/computing/raven-user-guide.html} {\enquote
  {\bibinfo {title} {{Raven User Guide}},}\ } (\bibinfo {year} {2023}),\
  \bibinfo {note} {[Online; accessed 01-June-2023; archived 01-June-2023;
  archive URL:
  https://web.archive.org/web/20230601200340/https://docs.mpcdf.mpg.de/doc/computing/raven-user-guide.html]}\BibitemShut
  {NoStop}%
\bibitem [{\citenamefont {Falgout}\ and\ \citenamefont
  {Yang}(2002)}]{falgout2002}%
  \BibitemOpen
  \bibfield  {author} {\bibinfo {author} {\bibfnamefont {R.~D.}\ \bibnamefont
  {Falgout}}\ and\ \bibinfo {author} {\bibfnamefont {U.~M.}\ \bibnamefont
  {Yang}},\ }in\ \href {\doibase 10.1007/3-540-47789-6_66} {\emph {\bibinfo
  {booktitle} {Computational Science --- ICCS 2002}}},\ \bibinfo {editor}
  {edited by\ \bibinfo {editor} {\bibfnamefont {P.~M.~A.}\ \bibnamefont
  {Sloot}}, \bibinfo {editor} {\bibfnamefont {A.~G.}\ \bibnamefont {Hoekstra}},
  \bibinfo {editor} {\bibfnamefont {C.~J.~K.}\ \bibnamefont {Tan}}, \ and\
  \bibinfo {editor} {\bibfnamefont {J.~J.}\ \bibnamefont {Dongarra}}}\
  (\bibinfo  {publisher} {Springer Berlin Heidelberg},\ \bibinfo {address}
  {Berlin, Heidelberg},\ \bibinfo {year} {2002})\ pp.\ \bibinfo {pages}
  {632--641}\BibitemShut {NoStop}%
\end{thebibliography}%

\end{document}